\documentclass[twocolumn,showpacs,floatfix,
superscriptaddress,
longbiblography,
amsmath,amssymb,
aps,
prl]{revtex4-2}

\usepackage{graphicx}
\usepackage{dcolumn}
\usepackage{amsmath}
\usepackage{bm}
\usepackage{appendix}
\usepackage{multirow}
\usepackage[bookmarks=true,colorlinks=true,urlcolor=blue,linkcolor=blue,citecolor=blue,breaklinks]{hyperref}
\usepackage[mathlines]{lineno}
\usepackage{bbold}
\usepackage{inconsolata} 
\usepackage[usenames,dvipsnames]{color} 
\usepackage{listings}
\usepackage{cancel}
\usepackage[normalem]{ulem}
\usepackage{pdfpages} 
\usepackage{pgffor} 

\makeatletter
\AtBeginDocument{\let\LS@rot\@undefined}
\makeatother

\def\supplementfilename{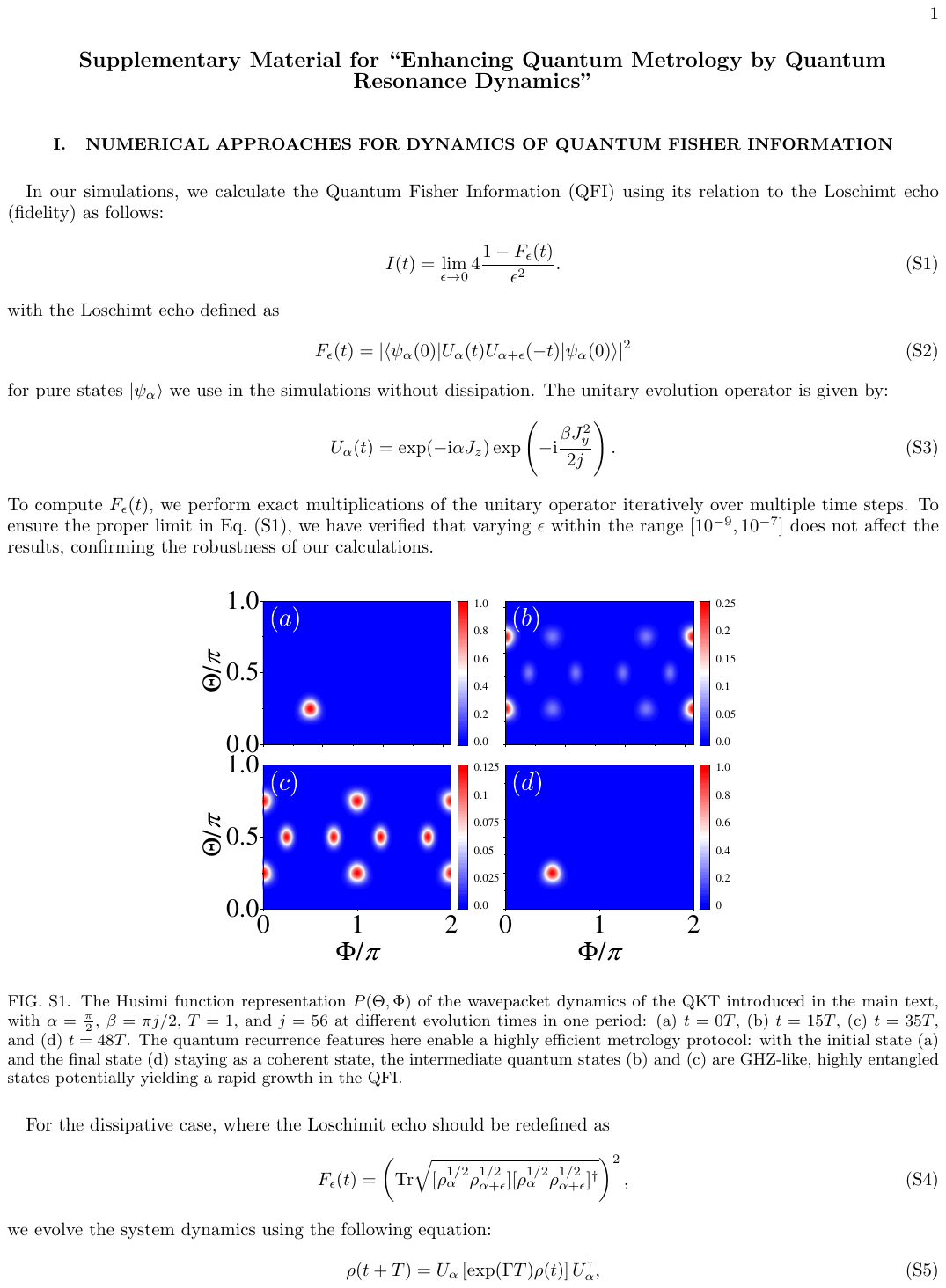}

\pdfximage{\supplementfilename}
\def\numbersupplementpages{\the\pdflastximagepages}

\newif\ifarXiv
\arXivtrue 

\renewcommand{\selectlanguage}[1]{}
\begin{document}

\preprint{APS/123-QED}

\title{Enhancing Quantum Metrology by Quantum Resonance Dynamics}

\author{Zhixing Zou}
\email{zouzx@nus.edu.sg}
\affiliation{Department of Physics, National University of Singapore, Singapore.}
\author{Jiangbin Gong}
\email{phygj@nus.edu.sg}
\affiliation{Department of Physics, National University of Singapore, Singapore.}
\affiliation{MajuLab, CNRS-UCA-SU-NUS-NTU International Joint Research Unit, Singapore.}
\affiliation{Centre for Quantum Technologies, National University of Singapore, Singapore.}

\author{Weitao Chen}
\email{chen.weitao@u.nus.edu}
\affiliation{Department of Physics, National University of Singapore, Singapore.}
\affiliation{MajuLab, CNRS-UCA-SU-NUS-NTU International Joint Research Unit, Singapore.}
\affiliation{Centre for Quantum Technologies, National University of Singapore, Singapore.}

\date{\today}

\begin{abstract}
Quantum effects in metrology can in principle enhance measurement precision from the so-called standard quantum limit to the Heisenberg Limit. Further advancements in quantum metrology largely rely on innovative metrology protocols that can avoid a number of known obstacles, including the challenge of preparing entangled states with sufficient fidelity, the readout noise in measuring highly entangled states, and no-go theorems for quantum metrology under noisy environments.  In this Letter,  exploiting some peculiar but experimentally feasible dynamical features of a collection of spins with all-to-all time-periodic interactions, we propose a metrology protocol that can circumvent all the three mentioned obstacles and yet still make good use of time as a resource for metrology.  Specifically,  by mapping the dynamics of such a periodically driven spin system to that of a paradigm of quantum chaos but tuned to some high-order quantum resonance, it is shown that a simple $SU(2)$ coherent state can, after evolving to highly entangled states in the ensuing dynamics, be dynamically brought back to the same initial coherent state.  The associated quantum Fisher information is found to exhibit quadratic scaling with both the number of spins and the duration of the metrology protocol. The achieved Heisenberg scaling can also largely survive in the presence of Markovian noise.   Representing a previously unknown strategy for quantum metrology,  the protocol proposed here can be tested on available experimental platforms.

\end{abstract}

\maketitle

\emph{Introduction.\textemdash}Quantum metrology has become an indispensable tool in advancing quantum technologies, driving breakthroughs in high-precision applications such as magnetometry \cite{Taylor2008,Brask2015,Danilin2018,Troiani2018}, gravitational wave detection \cite{Goda2008,Schnabel2010,Aasi2013,Vajente2019}, atomic clocks \cite{Andre2004,Katori2011,Borregaard2013,Kessler2014,Komar2014,Tse2019,Pezze2020,PedrozoPenafiel2020}, and navigation systems \cite{Giovannetti2001}. Classical noise imposes one constraint on a class of quantum metrology protocols, known as the standard quantum limit (SQL) such that measurement precision scales as $1/\sqrt{N}$ (with $N$ representing the number of qubits or atoms) \cite{Huelga1997,Escher2011,DemkowiczDobrzanski2012}. Surpassing the SQL requires to harness resourceful quantum effects without a classical analog.  The resulting Heisenberg Limit (HL) as a quantum precision limit scaling as $1/N$ epitomizes one big advantage of quantum metrology.

Achieving the HL experimentally remains a challenge despite substantial efforts \cite{Leibfried2004,Giovannetti2004,Giovannetti2006,DemkowiczDobrzanski2009,Kacprowicz2010,Giovannetti2011,Joo2011,Degen2017,Pezze2018,Daryanoosh2018,Barbieri2022,Huang2024}.  The preparation and readout of highly entangled states represent two major obstacles. Indeed, current proposals towards the HL typically rely on the Greenberger–Horne–Zeilinger (GHZ) state—a maximally entangled state \cite{Pezze2018} that can be notoriously difficult to prepare and protect in current experimental platforms \cite{Duer2002,Duer2004,Aolita2008,Lu2014,Wang2024}. The rich information content of GHZ states also makes them highly susceptible to readout noise \cite{Davis2017,Nolan2017}.  A third hurdle to overcome is the unavoidable noise experienced by a quantum metrology protocol, prompting some no-go theorems for noisy quantum metrology \cite{Braunstein1994,Huelga1997,Maccone2011,Smirne2016,ThomasPeter2011,Escher2011,Albarelli2018,Albarelli2022,Gorecki2022,Bai2023,Zhou2024}.  It is now known that environment noise of the Markovian type severely restricts information encoding time and hence makes the HL scaling difficult to achieve \cite{DemkowiczDobrzanski2009,Kacprowicz2010,DemkowiczDobrzanski2012}.  If a quantum metrology protocol is subject to such noise for too long,  its performance can even degrade below the SQL \cite{Huelga1997}.

\begin{figure}
\includegraphics[width=0.48\textwidth]{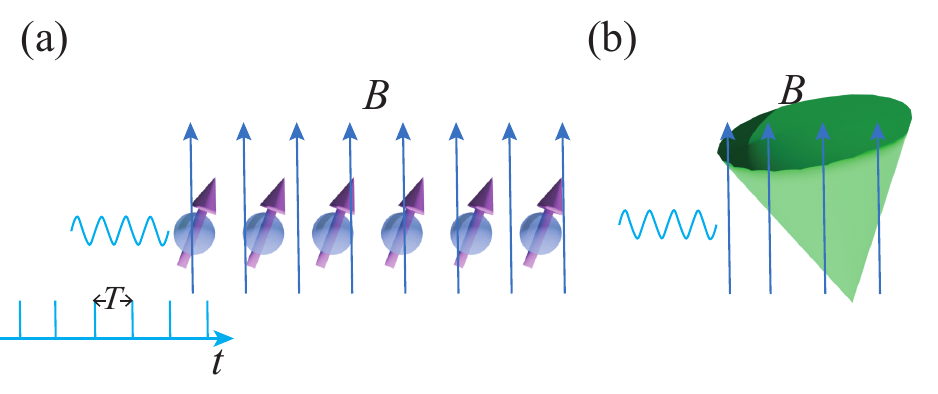}
    \caption{\label{fig1}Schematic of a quantum metrology protocol based on a system of $N$ spins subject to periodic modulation in their all-to-all interaction.  (a) A system of $N$ spins under a magnetic field, featuring all-to-all time-periodic interactions, where $N = 2j$. 
    (b) The setup of a quantum kicked top, describing the precession of a marco-spin of spin quantum number $j$ subject to a magnetic field along the $z$-axis and time-periodic twisting about the $y$-axis with period $T$. 
    }
\end{figure}

Based on the so-called quantum resonance dynamics \cite{Zou2022}, we advocate in this work an innovative metrology protocol that can go around all the three major issues mentioned above.  In a nutshell, the quantum resonance dynamics of a collection of spin systems subject to periodic all-to-all interaction can evolve, on its own, a simple easy-to-prepare initial state to a highly entangled state, and yet at the end, feature a complete recurrence of the initial state.  These dynamical features hence serve as a natural process to encode quantum information and also facilitate the quantum state readout at the end of the protocol. Furthermore, the quantum resonance dynamics featured below also allows us to analytically obtain the growth of the QFI with time, which is notably quadratic.   The quantum resonance dynamics used below hence can  exploit time very well as a resource for metrology. 

Specifically, we use spin coherent states with minimal quantum uncertainty as the initial state.  Such ``classical" initial states are of low-quantum-entanglement content \cite{Davis2017} to start with, hence experimentally friendly,  more robust against noise \cite{Davis2017,Huang2024} and also avoids readout noise 
upon quantum recurrence at the end of the metrology protocol. 
Furthermore, the time-periodic interaction applied to the spin system, designed to generate the quantum resonance dynamics, also places the quantum system under a useful non-equilibrium situation to effectively fight against decoherence \cite{Bai2023}.  With these advantages, we shall show in this work that the quantum Fisher information (QFI) of the adopted quantum resonance dynamics exhibits quadratic scaling with both the number of spins and the duration of the metrology protocol.  Even in the presence of Markovian noise,  the QFI can sustain a near-Heisenberg scaling in terms of the number of spins, though saturated with time.    
Representing a previously unknown strategy for quantum metrology,  we stress that  the protocol proposed here can be tested on available experimental platforms.

\emph{Quantum parameter estimation and dynamical sensitivity.\textemdash}Quantum parameter estimation theory focuses on the information extractable about an unknown parameter $\alpha$, which is encoded in a quantum state represented by the density operator $\rho_\alpha$. The quantitative measure of this information is the Quantum Fisher Information (QFI), which is directly related to the measurement precision through the well-known Cramér-Rao bound \cite{PARIS2009}. The QFI is defined as $I = \text{Tr}(\hat{L}_\alpha^2 \rho_\alpha)$, where $\hat{L}_\alpha$ is the symmetric logarithmic derivative satisfying $\partial_\alpha \rho_\alpha = (\hat{L}_\alpha \rho_\alpha + \rho_\alpha \hat{L}_\alpha)/2$ \cite{Liu2016}. Because the precision limit is proportional to $1/\sqrt{I}$ \cite{PARIS2009}, the QFI should scale as  $I \propto N$ to ahieve the SQL and $I \propto N^2$ to achieve the HL. On the other hand, a key measure of the dynamical sensitivity of a quantum system is the Loschmidt echo (fidelity) $F_{\epsilon}$.  The Loschmidt echo thus defined  quantifies the overlap between a state undergoing forward unitary evolution and the same state evolving backward under a slightly perturbed unitary operator \cite{Gorin2006}. Specifically, $F_{\epsilon}(t)=|\langle\psi_\alpha(0)|U_\alpha(t)U_{\alpha+\epsilon}(-t)|\psi_\alpha(0)\rangle|^2$ with $U_\alpha(t)$ the unitary operator of the evolution.   It can be shown that the QFI can be expressed in terms of the Loschmidt echo as $I(t)=\lim_{\epsilon\rightarrow0}4[1-F_\epsilon(t)]/\epsilon^2$ \cite{Braunstein1994,Fiderer2018}.  This makes it clear that the QFI is intrinsically linked to the dynamical sensitivity of quantum dynamics. 
For non-unitary evolution, the definition of the Loschmidt echo can be adjusted to $F_\epsilon(t)=||\rho_\alpha^{1/2}\rho_{\alpha+\epsilon}^{1/2}||_1^2$ where $ ||Q||_1\equiv\text{Tr}\sqrt{QQ^\dagger}$ \cite{Miszczak2009} and then a similar connection between dynamical sensitivity and the QFI can still be made. 

It is then evident that having the rapid growth of the QFI with encoding time is also pivotal for high-precision measurements. Because this growth is closely connected with the dynamical sensitivity of a quantum system,  previous researchers have gone beyond integrable models \cite{Pezze2018} and started to explore the potential of quantum chaos models for enhancing the growth of the QFI \cite{Fiderer2018}. Indeed, metrology protocols based on some quantum chaos models have the capability to build up quantum entanglement rapidly \cite{Fiderer2018,Lerose2020,Liu2021}.  This rapid growth of the quantum entanglement and hence the QFI can be linked with the underlying chaotic classical trajectories of the classical counterpart system, but only up to the Ehrenfest time \cite{Zaslavsky1981}.  Because the Ehrenfest time is typically logarithmically short \cite{Zaslavsky1981,Silvestrov2002}, the duration during which the QFI increases quickly is also rather limited.  This motivates us to seek alternative means to make good use of time as a precious resource.   In particular, the complete quantum recurrence in our metrology protocol below at particular timings indicates that the quantum dynamics simply repeats itself many times, and as such the growth of the QFI may be quadratic in terms of the sensing duration without saturation.

\begin{figure}
\includegraphics[width=0.5\textwidth]{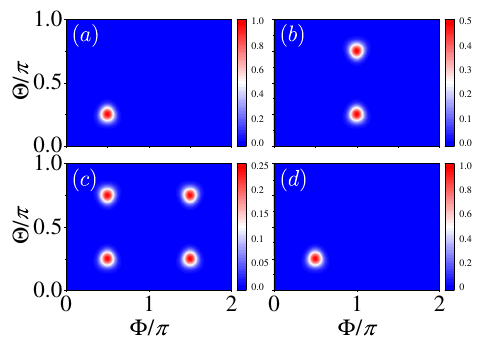}
    \caption{\label{fig2}
The Husimi function representation $P (\Theta, \Phi)$ of the wavepacket dynamics of the QKT introduced in the main text, with $\alpha = \frac{\pi}{2}$, $ \beta = \pi j$, $T = 1$, and $j = 56$ at different evolution times in one period: (a) $t = 0T$, (b) $t = 3T$, (c) $t =  6T$, and (d) $t = 8T$. The quantum recurrence features enable a highly efficient metrology protocol:  with the initial state (a) and the final state (d) staying as a coherent state, the intermediate quantum states (b) and (c) are GHZ-like, highly entangled states potentially yielding a rapid growth in the QFI.}
\end{figure}

\begin{figure}
\includegraphics[width=0.49\textwidth]{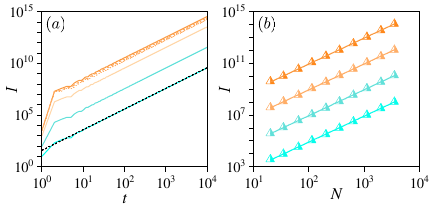}
    \caption{\label{fig3} Dynamics of the QFI associated with the quantum resonances of the QKT.  (a) The time evolution of the QFI. Cyan to orange curves represent different system sizes ranging from $N = 20$ to $N = 6324$, with $\beta = \pi j$. The dotted line corresponds to $\beta = \pi j + \delta$ with $\delta = 1.5$, and the dashed line represents $\delta = 2.0$. The black dashed line represents a power-law fit, showing $I \sim t^a$ with $a \approx 2$. (b) The dependence of the QFI on the number of spins for $\beta = \pi j$ at different times $t/T = 10, 10^2,10^3,10^4$ (from cyan to orange). The solid lines represent a fitting with $I \sim N^a$, yielding $a \approx 2$, demonstrating the Heisenberg  scaling}.
\end{figure}

\emph{A periodically driven spin system as a quantum sensor.\textemdash}
Consider an $N$-spin system with the following Hamiltonian \cite{Lerose2020}: 
\begin{equation}\small
    H_{\rm spin}=\sum_{k=1}^N \frac{\alpha}{2T}\sigma_{k}^z+\sum_{k=1}^N\sum_{l=1}^N\frac{\beta}{8j}\sigma_{k}^y\sigma_{l}^y \sum_{n=-\infty}^{\infty} \delta(t - nT),
\end{equation}
where $\sigma_{i}^{x,y,z}$ are the usual Pauli matrices for the spin indexed by $i$.  This Hamiltonian depicts a collection of spin-1/2 systems subject to periodic all-to-all interaction in the form of $\delta$-kicks with period $T$, in addition to a common field along the $x$ direction.  Here the external field parameter $\alpha$ could serve as the strength of a weak magnetic field to be probed.   We refer to Fig.~\ref{fig1} for a schematic illustration of this model. If we restrict the wave function to the symmetric-invariant subspace,  we may describe the quantum dynamics by introducing collective operators \cite{Lerose2020}.
By applying the transformation $J_\alpha = \sum_{i=1}^N (\sigma_{i}^\alpha/2)$ and assuming an initial collective state, the above Hamiltonian can be rewritten in the following form:
\begin{equation}
    H_{\text{QKT}}=\frac{\alpha J_z}{T}+\frac{\beta J_y^2}{2j} \sum_{n=-\infty}^{\infty} \delta(t - nT),
\end{equation}
 where $J_x$, $J_y$, and $J_z$ are collective angular momentum operators satisfying the commutation relations $[J_\alpha, J_\beta] = i \epsilon_{\alpha\beta\gamma} J_\gamma$, with $(J_x^2+J_y^2+J_z^2)|j,j_z\rangle=j(j+1)|j,j_z\rangle$  and $J_z |j, j_Z \rangle = j_z|j, j_z \rangle$, where $|j, j_z \rangle$ represents angular momentum eigenstates, and $j=N/2$.
 The unitary evolution operator of $H_{\text QKT}$ associated with one period $T$ is given by $U_{\text{QKT}} = \exp(-{\rm i} \alpha J_z) \exp\left[- {\rm i}(\beta J_y^2)/(2j)\right]$. Throughout we set $\hbar=1$.  Clearly then, our periodically kicked spin system as a quantum sensor is nothing but the celebrated quantum kicked top (QKT) model as one paradigm of quantum chaos \cite{Haake1987}.  Experimental realizations of the QKT are well-established, with implementations ranging from cold atoms \cite{Chaudhury2009} to trap ions \cite{Xu2010} and digital quantum simulators \cite{Sieberer2019}.


\begin{figure*}
\includegraphics[width=0.85\textwidth]{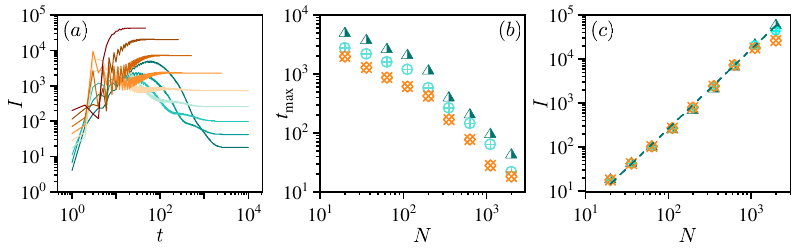}

    \caption{\label{fig4} The QFI in quantum resonance dynamics with dissipation. (a) Time evolution of the QFI in the quantum resonance dynamics of the QKT. Curves from cyan to orange represent different system sizes ranging from $N =20 $ to $N =2000 $, with $\beta = \pi j$.  (b) Dependence of $t_{\text max}$ on  the number of spins, at dissipation rates $\gamma = 0.0003, 0.0005, 0.0007$ (from dark green to red).  (c) Dependence of the saturated QFI on the number of spins for $\beta = \pi j$.  The dashed line represents a fit with $I \sim N^a$, yielding $a \approx 1.8$, which shows a near-Heisenberg  scaling. 
    System parameters are chosen to be the same as in panel (b) for different symbols.}
\end{figure*}

\emph{Achieving the HL via quantum resonance dynamics.\textemdash} Other than the correspondence between quantum dynamics and the underlying classical chaos, quantum chaos studies had uncovered other exotic aspects of quantum dynamics without a classical analog, with quantum resonance dynamics being one prominent example. Quantum resonance dynamics was first investigated in another quantum chaos model—the so-called quantum kicked rotor (QKR) model  \cite{Izrailev1980,Wimberger2003,Wimberger2004,Wimberger2005,Dana2006,Abb2009,McDowall2009,Billam2009,Talukdar2010,Tian2010,Sadgrove2011,Dubertrand2012,Ullah2012}, where energy absorption from the driving field may show ballistic growth albeit with corresponding classical trajectories being chaotic. The concept of quantum resonance does apply to the QKT, if the resonance condition $\beta = 4\pi jr/s$ is satisfied, with $r$ and $s$ being two coprime integers. Qualitatively, resonance arises due to the dynamical phase generated by the pulsed interaction. Specifically, the term 
$\exp\left[- {\rm i}(\beta J_y^2)/(2j)\right]$ generate dynamical phases with a special pattern (e.g., in the representation of the eigenstates of $J_y$) if the system parameter $\beta$ assumes values commensurate with $\pi$.    
Importantly, it is known that the quantum resonance dynamics of the QKT remains stable under a small perturbation $\delta$ to $\beta$. A full understanding the dynamics for general choices of $r$ and $s$, especially when they taking large integer values (hence very high order resonances) is challenging due to intricate number-theoretic properties, but a few special resonance cases were studied \cite{Zou2022,Anand2024} previously. Fortunately, for our purpose here, certain special resonance cases already suffice to design useful metrology protocols.

The quantum resonance dynamics of the QKT can even exhibit complete quantum recurrence for some specific system parameters under the condition of $\alpha = \pi/2$: (i) For $\beta = 2\pi j$, we have $U_{2\pi j}^2 = I$, and so the dynamics are periodic with a period of $2T$. However, such dynamics generates no quantum entanglement during the time evolution. Indeed, a simple examination shows that the associated one-step unitary operator is composed of local unitaries only. From quantum metrology perspective, this dynamics is not resourceful since it cannot be used to encode quantum information with dynamics-generated entanglement. (ii) For $\beta = \pi j$, we have $U_{\pi j}^8 = I$. In this case the dynamics exhibits complete quantum recurrence with a period of $8T$. This case is drastically different because it generates several GHZ-like states during the middle steps of a complete recurrence process, thus involving entanglement heavily. Due to this key observation,  we may simply pick the $SU(2)$ coherent state with minimum quantum certainty as the easy-to-prepare initial state, whereas the final state is guaranteed to be the same coherent state due to the perfect quantum recurrence. We thereby can exploit this case  to facilitate straightforward state preparation and readout processes for quantum metrology. (iii) For $\beta = \pi j/2$, numerical studies indicate that $U_{\pi j/2}^{48} = I$, showing perfect recurrence behavior similar to case (ii) but with a longer recurrence time while generating highly entangled states during the dynamical process.  This makes case (iii) another promising parameter choice for quantum metrology. In this Letter, we focus on the case (ii) with $\beta = \pi j$ to illustrate our central ideas.

We have performed computer simulations starting from an initial coherent state $|\Theta, \Phi \rangle = \exp\left[{\rm i} \Theta (J_x \sin\Phi - J_y \cos\Phi)\right] |j, j \rangle$. The Husimi representation $P(\Theta, \Phi) = \langle \Theta, \Phi |\rho|\Theta, \Phi \rangle$, where $\rho$ is the density matrix, is used to visualize the wave packet dynamics \cite{Agarwal1981}. The results are presented in Fig.~\ref{fig2}, with additional simulations for $\beta = \pi j/2$ in Supplementary Material~\cite{sup}.   As shown in  Fig.~\ref{fig2}, the initial coherent state is split into a superposition of two well-separated wavepackets at $t=3T$ and a superposition of four well-separated wavepckets at $t=6T$, before a perfect quantum recurrence occurs at $t=8T$. These superposition states represent highly entangled states involving all the $N$ spins.  Because the dynamical sensitivity is really about the sensitivity of one whole process, the generation of entangled states in the middle steps of a dynamical process suggests the possibility of Heisenberg scaling of the QFI, an insight that is indeed confirmed later.  

One main difference between the resonance dynamics illustrated above and the classical-like chaotic wavepacket dynamics is that quantum resonance dynamics is found to exhibit the so-called  \textit{pseudo-classical stability}, a stability feature that can only be well explained by another effective classical system after technical treatments of quantum resonances.  Without presenting such technical details, we highlight that unlike typical diffusive wavepacket dynamics resembling that of an ensemble of classical chaotic trajectories,  quantum resonance dynamics can be connected with stable pseudo-classical trajectories, and as such we can obtain stable time-evolving wavepackets: they are not spreading with irregular patterns but instead split into multiple well-separated wavepackets during the time evolution.  This ensures that the resonance dynamics remain controlled and conducive to generating a rapid increase in the QFI throughout the time evolution \cite{Zou2022}.  Indeed, leveraging the solvable nature of quantum resonance dynamics, we have analytically obtained that the scaling of the QFI is given by $I(t) \propto t^2$ (see Supplementary Material~\cite{sup}). Computer simulations further confirm this prediction of the QFI growth. The simulation results, shown in Fig.~\ref{fig3}(a), reveal that this quadratic scaling remains robust under small perturbations $\delta$ to the value of $\beta$, thus making our protocol experimental feasible. In contrast to the Ehrenfest time limitation imposed on all sensors based on fast wavepacket spreading, the unbounded long-time quadratic growth of QFI under quantum resonance dynamics is also confirmed in Fig.~\ref{fig3}(a). 

Computer simulations of the quantum dynamics with varying system sizes also confirm a perfect Heisenberg Limit (HL) scaling, namely, $I(N) \propto N^2$, as shown in Fig.~\ref{fig3}(b). As conjectured above, this desirable scaling can be connected with the generation of highly entangled states at times other than perfect quantum recurrence times, as illustrated in Fig.~\ref{fig1}. Remarkably, if we now examine the dynamical sensitivity or the QFI over many periods of quantum recurrence, we find that this Heisenberg scaling persists, and hence in terms of both time and the number of spins, we have $I_{\text{QKT}} \propto t^2 N^2$. This combination of advantageous scaling for an initial coherent state really highlights the power of quantum resonance dynamics in enhancing quantum metrology.  Indeed, as a comparison, the chaotic kicked top dynamics can only generate $t^2N$ scaling of the QFI, up to the Heisenberg time, followed by a scaling behavior of $tN^2$ scaling due to the saturation of the spreading of the time evolving state ~\cite{Fiderer2018}.

\emph{Noisy quantum metrology.\textemdash}Dissipative noise remains a significant challenge in modern quantum metrology experiments by limiting the information encoding time. Existing no-go theorems for noisy quantum metrology \cite{{Braunstein1994,Huelga1997,Maccone2011,Smirne2016,ThomasPeter2011,Escher2011,Albarelli2018,Albarelli2022,Gorecki2022,Bai2023,Zhou2024}} best illustrate this point. However, because the time-periodic interaction to engineer the quantum resonance dynamics also places our quantum system under a non-equilibrium condition, previous no-go theorems may not apply \cite{Bai2023}.  We thus examine the dissipative dynamics of the QKT under Markovian noise to examine the possible quantum advantages of our metrology protocol under an adverse environment. The quantum resonance dynamics under Markovian noise can be modeled by the Master equation $\dot{\rho}(t)= -i[H,\rho(t)]+\mathcal{D}[\rho(t)]$. Here we consider superradiant damping \cite{Bonifacio1971,Gross1982}, which allows simulations for large system sizes up to $N = 2\times10^3$. The superradiant damping can be expressed analytically as $\mathcal{D}[\rho(t)] = \gamma\left( [J_{-}, \rho(t) J_{+}]+ [J_{-}\rho(t),  J_{+}] \right)\equiv\Gamma \rho(t)$  with $J_{\pm}=J_x\pm iJ_y$ and $\gamma$ denoting the dissipation strength \cite{Haake2018}. Consequently, the evolution of the density operator from $t$ to $t+T$ is given by:
\begin{equation}
    \rho(t+T)=U\left[\exp(\Gamma T)\rho(t)\right]U^\dagger.
\end{equation} 
We present in Fig.~\ref{fig4}(a) our simulation results of the QFI dynamics, again starting from a coherent state.  As expected, a saturation of the QFI is observed after a certain time $t_{\rm max}$. This time threshold $t_{\rm max}$ certainly reflects a balance between dynamics-generated quantum features and decoherence. Fig.~\ref{fig4}(b) presents a quantitative relation between $t_{\rm max}$ and the number of spins.  Of more interest is the scaling of the saturated QFI with the number of spins.  Notably, we find a near-Heisenberg scaling with $I \propto N^a$, where $a \rightarrow 1.8$, as shown in Fig.~\ref{fig4}(c). Such near-HL scaling, achieved with an initial coherent state in a noisy environment, represents significant progress as compared to many other metrology protocols that require the use of a GHZ state as the initial state. The use of quantum resonance dynamics is hence useful in enhancing quantum metrology even in the presence of an environment.  

\emph{Conclusion.\textemdash}We have shown a number of advantages in using one class of quantum resonance dynamics for quantum metrology. 
 The approach advocated in this work has capabilities to overcome current key challenges in pushing the precision limit of quantum metrology, including the inefficient QFI scaling of initial coherent states under unstable time evolution, the difficulty in preparing GHZ-like states, the limitations imposed by existing no-go theorems in noisy metrology, and the readout noise of highly entangled states emerging at the end of a metrology protocol.    The quantum recurrence intrinsic to the quantum resonance dynamics examined here enables highly efficient encoding and readout procedures.  As a result, the QFI of even analytically solvable quantum resonance dynamics  exhibits the Heisenberg scaling with the number of spins. Furthermore, both numerical and analytical investigations for closed systems show that the growth of the QFI can be quadratic in time without a bound,  even for times far beyond the Heisenberg time.  For the quantum resonance dynamics with dissipation, thanks to the non-equilibrium nature of our system, near-Heisenberg scaling persists.  This work is of immediate experimental interest because we only need to place the already experimentally realized dynamical model in a parameter regime different than before.  In the future we plan to extend our approach to other settings, such as atomic spin-precession magnetometers \cite{Allred2002,Kominis2003,Savukov2005,Budker2007,Sheng2013}.

\acknowledgements
Z.Z. and J.G. are  supported by the National Research Foundation, Singapore, under its Competitive Research Programme (CRP Award No: NRF-CRP30-2023-0002). Any opinions, findings and conclusions or recommendations expressed in this material are those of the author(s) and do not reflect the views of National Research Foundation, Singapore. 
W.~Chen wishes to thank G.~Lemari\'e for mentorship in numerical simulations on related projects and B.~Georgeot for assistance in acquiring computational resources. We thank Calcul en Midi-Pyrénées
(CALMIP) for computational resources and assistance. W.~Chen is supported by the President's Graduate Fellowship at National University of Singapore and the Merlion Ph.D. Scholarship awarded by the French Embassy in Singapore. 


\bibliography{first_submission/apssamp}

\providecommand{\noopsort}[1]{}\providecommand{\singleletter}[1]{#1}%
\begin{thebibliography}{86}%
\makeatletter
\providecommand \@ifxundefined [1]{%
 \@ifx{#1\undefined}
}%
\providecommand \@ifnum [1]{%
 \ifnum #1\expandafter \@firstoftwo
 \else \expandafter \@secondoftwo
 \fi
}%
\providecommand \@ifx [1]{%
 \ifx #1\expandafter \@firstoftwo
 \else \expandafter \@secondoftwo
 \fi
}%
\providecommand \natexlab [1]{#1}%
\providecommand \enquote  [1]{``#1''}%
\providecommand \bibnamefont  [1]{#1}%
\providecommand \bibfnamefont [1]{#1}%
\providecommand \citenamefont [1]{#1}%
\providecommand \href@noop [0]{\@secondoftwo}%
\providecommand \href [0]{\begingroup \@sanitize@url \@href}%
\providecommand \@href[1]{\@@startlink{#1}\@@href}%
\providecommand \@@href[1]{\endgroup#1\@@endlink}%
\providecommand \@sanitize@url [0]{\catcode `\\12\catcode `\$12\catcode `\&12\catcode `\#12\catcode `\^12\catcode `\_12\catcode `\%12\relax}%
\providecommand \@@startlink[1]{}%
\providecommand \@@endlink[0]{}%
\providecommand \url  [0]{\begingroup\@sanitize@url \@url }%
\providecommand \@url [1]{\endgroup\@href {#1}{\urlprefix }}%
\providecommand \urlprefix  [0]{URL }%
\providecommand \Eprint [0]{\href }%
\providecommand \doibase [0]{https://doi.org/}%
\providecommand \selectlanguage [0]{\@gobble}%
\providecommand \bibinfo  [0]{\@secondoftwo}%
\providecommand \bibfield  [0]{\@secondoftwo}%
\providecommand \translation [1]{[#1]}%
\providecommand \BibitemOpen [0]{}%
\providecommand \bibitemStop [0]{}%
\providecommand \bibitemNoStop [0]{.\EOS\space}%
\providecommand \EOS [0]{\spacefactor3000\relax}%
\providecommand \BibitemShut  [1]{\csname bibitem#1\endcsname}%
\let\auto@bib@innerbib\@empty
\bibitem [{\citenamefont {Taylor}\ \emph {et~al.}(2008)\citenamefont {Taylor}, \citenamefont {Cappellaro}, \citenamefont {Childress}, \citenamefont {Jiang}, \citenamefont {Budker}, \citenamefont {Hemmer}, \citenamefont {Yacoby}, \citenamefont {Walsworth},\ and\ \citenamefont {Lukin}}]{Taylor2008}%
  \BibitemOpen
  \bibfield  {author} {\bibinfo {author} {\bibfnamefont {J.~M.}\ \bibnamefont {Taylor}}, \bibinfo {author} {\bibfnamefont {P.}~\bibnamefont {Cappellaro}}, \bibinfo {author} {\bibfnamefont {L.}~\bibnamefont {Childress}}, \bibinfo {author} {\bibfnamefont {L.}~\bibnamefont {Jiang}}, \bibinfo {author} {\bibfnamefont {D.}~\bibnamefont {Budker}}, \bibinfo {author} {\bibfnamefont {P.~R.}\ \bibnamefont {Hemmer}}, \bibinfo {author} {\bibfnamefont {A.}~\bibnamefont {Yacoby}}, \bibinfo {author} {\bibfnamefont {R.}~\bibnamefont {Walsworth}},\ and\ \bibinfo {author} {\bibfnamefont {M.~D.}\ \bibnamefont {Lukin}},\ }\bibfield  {title} {\bibinfo {title} {High-sensitivity diamond magnetometer with nanoscale resolution},\ }\href {https://doi.org/10.1038/nphys1075} {\bibfield  {journal} {\bibinfo  {journal} {Nat. Phys.}\ }\textbf {\bibinfo {volume} {4}},\ \bibinfo {pages} {810} (\bibinfo {year} {2008})}\BibitemShut {NoStop}%
\bibitem [{\citenamefont {Brask}\ \emph {et~al.}(2015)\citenamefont {Brask}, \citenamefont {Chaves},\ and\ \citenamefont {Ko\l{}ody\ifmmode~\acute{n}\else \'{n}\fi{}ski}}]{Brask2015}%
  \BibitemOpen
  \bibfield  {author} {\bibinfo {author} {\bibfnamefont {J.~B.}\ \bibnamefont {Brask}}, \bibinfo {author} {\bibfnamefont {R.}~\bibnamefont {Chaves}},\ and\ \bibinfo {author} {\bibfnamefont {J.}~\bibnamefont {Ko\l{}ody\ifmmode~\acute{n}\else \'{n}\fi{}ski}},\ }\bibfield  {title} {\bibinfo {title} {Improved quantum magnetometry beyond the standard quantum limit},\ }\href {https://doi.org/10.1103/PhysRevX.5.031010} {\bibfield  {journal} {\bibinfo  {journal} {Phys. Rev. X}\ }\textbf {\bibinfo {volume} {5}},\ \bibinfo {pages} {031010} (\bibinfo {year} {2015})}\BibitemShut {NoStop}%
\bibitem [{\citenamefont {Danilin}\ \emph {et~al.}(2018)\citenamefont {Danilin}, \citenamefont {Lebedev}, \citenamefont {Vepsäläinen}, \citenamefont {Lesovik}, \citenamefont {Blatter},\ and\ \citenamefont {Paraoanu}}]{Danilin2018}%
  \BibitemOpen
  \bibfield  {author} {\bibinfo {author} {\bibfnamefont {S.}~\bibnamefont {Danilin}}, \bibinfo {author} {\bibfnamefont {A.~V.}\ \bibnamefont {Lebedev}}, \bibinfo {author} {\bibfnamefont {A.}~\bibnamefont {Vepsäläinen}}, \bibinfo {author} {\bibfnamefont {G.~B.}\ \bibnamefont {Lesovik}}, \bibinfo {author} {\bibfnamefont {G.}~\bibnamefont {Blatter}},\ and\ \bibinfo {author} {\bibfnamefont {G.~S.}\ \bibnamefont {Paraoanu}},\ }\bibfield  {title} {\bibinfo {title} {Quantum-enhanced magnetometry by phase estimation algorithms with a single artificial atom},\ }\href {https://doi.org/10.1038/s41534-018-0078-y} {\bibfield  {journal} {\bibinfo  {journal} {npj Quantum Inf.}\ }\textbf {\bibinfo {volume} {4}},\ \bibinfo {pages} {29} (\bibinfo {year} {2018})}\BibitemShut {NoStop}%
\bibitem [{\citenamefont {Troiani}\ and\ \citenamefont {Paris}(2018)}]{Troiani2018}%
  \BibitemOpen
  \bibfield  {author} {\bibinfo {author} {\bibfnamefont {F.}~\bibnamefont {Troiani}}\ and\ \bibinfo {author} {\bibfnamefont {M.~G.~A.}\ \bibnamefont {Paris}},\ }\bibfield  {title} {\bibinfo {title} {Universal quantum magnetometry with spin states at equilibrium},\ }\href {https://doi.org/10.1103/PhysRevLett.120.260503} {\bibfield  {journal} {\bibinfo  {journal} {Phys. Rev. Lett.}\ }\textbf {\bibinfo {volume} {120}},\ \bibinfo {pages} {260503} (\bibinfo {year} {2018})}\BibitemShut {NoStop}%
\bibitem [{\citenamefont {Goda}\ \emph {et~al.}(2008)\citenamefont {Goda}, \citenamefont {Miyakawa}, \citenamefont {Mikhailov}, \citenamefont {Saraf}, \citenamefont {Adhikari}, \citenamefont {McKenzie}, \citenamefont {Ward}, \citenamefont {Vass}, \citenamefont {Weinstein},\ and\ \citenamefont {Mavalvala}}]{Goda2008}%
  \BibitemOpen
  \bibfield  {author} {\bibinfo {author} {\bibfnamefont {K.}~\bibnamefont {Goda}}, \bibinfo {author} {\bibfnamefont {O.}~\bibnamefont {Miyakawa}}, \bibinfo {author} {\bibfnamefont {E.~E.}\ \bibnamefont {Mikhailov}}, \bibinfo {author} {\bibfnamefont {S.}~\bibnamefont {Saraf}}, \bibinfo {author} {\bibfnamefont {R.}~\bibnamefont {Adhikari}}, \bibinfo {author} {\bibfnamefont {K.}~\bibnamefont {McKenzie}}, \bibinfo {author} {\bibfnamefont {R.}~\bibnamefont {Ward}}, \bibinfo {author} {\bibfnamefont {S.}~\bibnamefont {Vass}}, \bibinfo {author} {\bibfnamefont {A.~J.}\ \bibnamefont {Weinstein}},\ and\ \bibinfo {author} {\bibfnamefont {N.}~\bibnamefont {Mavalvala}},\ }\bibfield  {title} {\bibinfo {title} {A quantum-enhanced prototype gravitational-wave detector},\ }\href {https://doi.org/10.1038/nphys920} {\bibfield  {journal} {\bibinfo  {journal} {Nat. Phys.}\ }\textbf {\bibinfo {volume} {4}},\ \bibinfo {pages} {472} (\bibinfo {year} {2008})}\BibitemShut {NoStop}%
\bibitem [{\citenamefont {Schnabel}\ \emph {et~al.}(2010)\citenamefont {Schnabel}, \citenamefont {Mavalvala}, \citenamefont {McClelland},\ and\ \citenamefont {Lam}}]{Schnabel2010}%
  \BibitemOpen
  \bibfield  {author} {\bibinfo {author} {\bibfnamefont {R.}~\bibnamefont {Schnabel}}, \bibinfo {author} {\bibfnamefont {N.}~\bibnamefont {Mavalvala}}, \bibinfo {author} {\bibfnamefont {D.~E.}\ \bibnamefont {McClelland}},\ and\ \bibinfo {author} {\bibfnamefont {P.~K.}\ \bibnamefont {Lam}},\ }\bibfield  {title} {\bibinfo {title} {Quantum metrology for gravitational wave astronomy},\ }\href {https://doi.org/10.1038/ncomms1122} {\bibfield  {journal} {\bibinfo  {journal} {Nat. Commun.}\ }\textbf {\bibinfo {volume} {1}},\ \bibinfo {pages} {121} (\bibinfo {year} {2010})}\BibitemShut {NoStop}%
\bibitem [{\citenamefont {Aasi}\ \emph {et~al.}(2013)\citenamefont {Aasi} \emph {et~al.}}]{Aasi2013}%
  \BibitemOpen
  \bibfield  {author} {\bibinfo {author} {\bibfnamefont {J.}~\bibnamefont {Aasi}} \emph {et~al.},\ }\bibfield  {title} {\bibinfo {title} {Enhanced sensitivity of the ligo gravitational wave detector by using squeezed states of light},\ }\href {https://doi.org/10.1038/nphoton.2013.177} {\bibfield  {journal} {\bibinfo  {journal} {Nat. Photonics}\ }\textbf {\bibinfo {volume} {7}},\ \bibinfo {pages} {613} (\bibinfo {year} {2013})}\BibitemShut {NoStop}%
\bibitem [{\citenamefont {Vajente}\ \emph {et~al.}(2019)\citenamefont {Vajente}, \citenamefont {Gustafson},\ and\ \citenamefont {Reitze}}]{Vajente2019}%
  \BibitemOpen
  \bibfield  {author} {\bibinfo {author} {\bibfnamefont {G.}~\bibnamefont {Vajente}}, \bibinfo {author} {\bibfnamefont {E.~K.}\ \bibnamefont {Gustafson}},\ and\ \bibinfo {author} {\bibfnamefont {D.~H.}\ \bibnamefont {Reitze}},\ }\bibfield  {title} {\bibinfo {title} {Precision interferometry for gravitational wave detection: Current status and future trends}\ }(\bibinfo  {publisher} {Academic Press},\ \bibinfo {year} {2019})\ pp.\ \bibinfo {pages} {75--148}\BibitemShut {NoStop}%
\bibitem [{\citenamefont {Andr\'e}\ \emph {et~al.}(2004)\citenamefont {Andr\'e}, \citenamefont {S\o{}rensen},\ and\ \citenamefont {Lukin}}]{Andre2004}%
  \BibitemOpen
  \bibfield  {author} {\bibinfo {author} {\bibfnamefont {A.}~\bibnamefont {Andr\'e}}, \bibinfo {author} {\bibfnamefont {A.~S.}\ \bibnamefont {S\o{}rensen}},\ and\ \bibinfo {author} {\bibfnamefont {M.~D.}\ \bibnamefont {Lukin}},\ }\bibfield  {title} {\bibinfo {title} {Stability of atomic clocks based on entangled atoms},\ }\href {https://doi.org/10.1103/PhysRevLett.92.230801} {\bibfield  {journal} {\bibinfo  {journal} {Phys. Rev. Lett.}\ }\textbf {\bibinfo {volume} {92}},\ \bibinfo {pages} {230801} (\bibinfo {year} {2004})}\BibitemShut {NoStop}%
\bibitem [{\citenamefont {Katori}(2011)}]{Katori2011}%
  \BibitemOpen
  \bibfield  {author} {\bibinfo {author} {\bibfnamefont {H.}~\bibnamefont {Katori}},\ }\bibfield  {title} {\bibinfo {title} {Optical lattice clocks and quantum metrology},\ }\href {https://doi.org/10.1038/nphoton.2011.45} {\bibfield  {journal} {\bibinfo  {journal} {Nat. Photonics}\ }\textbf {\bibinfo {volume} {5}},\ \bibinfo {pages} {203} (\bibinfo {year} {2011})}\BibitemShut {NoStop}%
\bibitem [{\citenamefont {Borregaard}\ and\ \citenamefont {S\o{}rensen}(2013)}]{Borregaard2013}%
  \BibitemOpen
  \bibfield  {author} {\bibinfo {author} {\bibfnamefont {J.}~\bibnamefont {Borregaard}}\ and\ \bibinfo {author} {\bibfnamefont {A.~S.}\ \bibnamefont {S\o{}rensen}},\ }\bibfield  {title} {\bibinfo {title} {Near-heisenberg-limited atomic clocks in the presence of decoherence},\ }\href {https://doi.org/10.1103/PhysRevLett.111.090801} {\bibfield  {journal} {\bibinfo  {journal} {Phys. Rev. Lett.}\ }\textbf {\bibinfo {volume} {111}},\ \bibinfo {pages} {090801} (\bibinfo {year} {2013})}\BibitemShut {NoStop}%
\bibitem [{\citenamefont {Kessler}\ \emph {et~al.}(2014)\citenamefont {Kessler}, \citenamefont {K\'om\'ar}, \citenamefont {Bishof}, \citenamefont {Jiang}, \citenamefont {S\o{}rensen}, \citenamefont {Ye},\ and\ \citenamefont {Lukin}}]{Kessler2014}%
  \BibitemOpen
  \bibfield  {author} {\bibinfo {author} {\bibfnamefont {E.~M.}\ \bibnamefont {Kessler}}, \bibinfo {author} {\bibfnamefont {P.}~\bibnamefont {K\'om\'ar}}, \bibinfo {author} {\bibfnamefont {M.}~\bibnamefont {Bishof}}, \bibinfo {author} {\bibfnamefont {L.}~\bibnamefont {Jiang}}, \bibinfo {author} {\bibfnamefont {A.~S.}\ \bibnamefont {S\o{}rensen}}, \bibinfo {author} {\bibfnamefont {J.}~\bibnamefont {Ye}},\ and\ \bibinfo {author} {\bibfnamefont {M.~D.}\ \bibnamefont {Lukin}},\ }\bibfield  {title} {\bibinfo {title} {Heisenberg-limited atom clocks based on entangled qubits},\ }\href {https://doi.org/10.1103/PhysRevLett.112.190403} {\bibfield  {journal} {\bibinfo  {journal} {Phys. Rev. Lett.}\ }\textbf {\bibinfo {volume} {112}},\ \bibinfo {pages} {190403} (\bibinfo {year} {2014})}\BibitemShut {NoStop}%
\bibitem [{\citenamefont {Kómár}\ \emph {et~al.}(2014)\citenamefont {Kómár}, \citenamefont {Kessler}, \citenamefont {Bishof}, \citenamefont {Jiang}, \citenamefont {Sørensen}, \citenamefont {Ye},\ and\ \citenamefont {Lukin}}]{Komar2014}%
  \BibitemOpen
  \bibfield  {author} {\bibinfo {author} {\bibfnamefont {P.}~\bibnamefont {Kómár}}, \bibinfo {author} {\bibfnamefont {E.~M.}\ \bibnamefont {Kessler}}, \bibinfo {author} {\bibfnamefont {M.}~\bibnamefont {Bishof}}, \bibinfo {author} {\bibfnamefont {L.}~\bibnamefont {Jiang}}, \bibinfo {author} {\bibfnamefont {A.~S.}\ \bibnamefont {Sørensen}}, \bibinfo {author} {\bibfnamefont {J.}~\bibnamefont {Ye}},\ and\ \bibinfo {author} {\bibfnamefont {M.~D.}\ \bibnamefont {Lukin}},\ }\bibfield  {title} {\bibinfo {title} {A quantum network of clocks},\ }\href {https://doi.org/10.1038/nphys3000} {\bibfield  {journal} {\bibinfo  {journal} {Nat. Phys.}\ }\textbf {\bibinfo {volume} {10}},\ \bibinfo {pages} {582} (\bibinfo {year} {2014})}\BibitemShut {NoStop}%
\bibitem [{\citenamefont {Tse}(2019)}]{Tse2019}%
  \BibitemOpen
  \bibfield  {author} {\bibinfo {author} {\bibfnamefont {M.~i.}\ \bibnamefont {Tse}},\ }\bibfield  {title} {\bibinfo {title} {Quantum-enhanced advanced ligo detectors in the era of gravitational-wave astronomy},\ }\href {https://doi.org/10.1103/PhysRevLett.123.231107} {\bibfield  {journal} {\bibinfo  {journal} {Phys. Rev. Lett.}\ }\textbf {\bibinfo {volume} {123}},\ \bibinfo {pages} {231107} (\bibinfo {year} {2019})}\BibitemShut {NoStop}%
\bibitem [{\citenamefont {Pezz\`e}\ and\ \citenamefont {Smerzi}(2020)}]{Pezze2020}%
  \BibitemOpen
  \bibfield  {author} {\bibinfo {author} {\bibfnamefont {L.}~\bibnamefont {Pezz\`e}}\ and\ \bibinfo {author} {\bibfnamefont {A.}~\bibnamefont {Smerzi}},\ }\bibfield  {title} {\bibinfo {title} {Heisenberg-limited noisy atomic clock using a hybrid coherent and squeezed state protocol},\ }\href {https://doi.org/10.1103/PhysRevLett.125.210503} {\bibfield  {journal} {\bibinfo  {journal} {Phys. Rev. Lett.}\ }\textbf {\bibinfo {volume} {125}},\ \bibinfo {pages} {210503} (\bibinfo {year} {2020})}\BibitemShut {NoStop}%
\bibitem [{\citenamefont {Pedrozo-Peñafiel}\ \emph {et~al.}(2020)\citenamefont {Pedrozo-Peñafiel}, \citenamefont {Colombo}, \citenamefont {Shu}, \citenamefont {Adiyatullin}, \citenamefont {Li}, \citenamefont {Mendez}, \citenamefont {Braverman}, \citenamefont {Kawasaki}, \citenamefont {Akamatsu}, \citenamefont {Xiao},\ and\ \citenamefont {Vuletić}}]{PedrozoPenafiel2020}%
  \BibitemOpen
  \bibfield  {author} {\bibinfo {author} {\bibfnamefont {E.}~\bibnamefont {Pedrozo-Peñafiel}}, \bibinfo {author} {\bibfnamefont {S.}~\bibnamefont {Colombo}}, \bibinfo {author} {\bibfnamefont {C.}~\bibnamefont {Shu}}, \bibinfo {author} {\bibfnamefont {A.~F.}\ \bibnamefont {Adiyatullin}}, \bibinfo {author} {\bibfnamefont {Z.}~\bibnamefont {Li}}, \bibinfo {author} {\bibfnamefont {E.}~\bibnamefont {Mendez}}, \bibinfo {author} {\bibfnamefont {B.}~\bibnamefont {Braverman}}, \bibinfo {author} {\bibfnamefont {A.}~\bibnamefont {Kawasaki}}, \bibinfo {author} {\bibfnamefont {D.}~\bibnamefont {Akamatsu}}, \bibinfo {author} {\bibfnamefont {Y.}~\bibnamefont {Xiao}},\ and\ \bibinfo {author} {\bibfnamefont {V.}~\bibnamefont {Vuletić}},\ }\bibfield  {title} {\bibinfo {title} {Entanglement on an optical atomic-clock transition},\ }\href {https://doi.org/10.1038/s41586-020-3006-1} {\bibfield  {journal} {\bibinfo  {journal} {Nature}\ }\textbf {\bibinfo {volume} {588}},\ \bibinfo {pages} {414} (\bibinfo {year}
  {2020})}\BibitemShut {NoStop}%
\bibitem [{\citenamefont {Giovannetti}\ \emph {et~al.}(2001)\citenamefont {Giovannetti}, \citenamefont {Lloyd},\ and\ \citenamefont {Maccone}}]{Giovannetti2001}%
  \BibitemOpen
  \bibfield  {author} {\bibinfo {author} {\bibfnamefont {V.}~\bibnamefont {Giovannetti}}, \bibinfo {author} {\bibfnamefont {S.}~\bibnamefont {Lloyd}},\ and\ \bibinfo {author} {\bibfnamefont {L.}~\bibnamefont {Maccone}},\ }\bibfield  {title} {\bibinfo {title} {Quantum-enhanced positioning and clock synchronization},\ }\href {https://doi.org/10.1038/35086525} {\bibfield  {journal} {\bibinfo  {journal} {Nature}\ }\textbf {\bibinfo {volume} {412}},\ \bibinfo {pages} {417} (\bibinfo {year} {2001})}\BibitemShut {NoStop}%
\bibitem [{\citenamefont {Huelga}\ \emph {et~al.}(1997)\citenamefont {Huelga}, \citenamefont {Macchiavello}, \citenamefont {Pellizzari}, \citenamefont {Ekert}, \citenamefont {Plenio},\ and\ \citenamefont {Cirac}}]{Huelga1997}%
  \BibitemOpen
  \bibfield  {author} {\bibinfo {author} {\bibfnamefont {S.~F.}\ \bibnamefont {Huelga}}, \bibinfo {author} {\bibfnamefont {C.}~\bibnamefont {Macchiavello}}, \bibinfo {author} {\bibfnamefont {T.}~\bibnamefont {Pellizzari}}, \bibinfo {author} {\bibfnamefont {A.~K.}\ \bibnamefont {Ekert}}, \bibinfo {author} {\bibfnamefont {M.~B.}\ \bibnamefont {Plenio}},\ and\ \bibinfo {author} {\bibfnamefont {J.~I.}\ \bibnamefont {Cirac}},\ }\bibfield  {title} {\bibinfo {title} {Improvement of frequency standards with quantum entanglement},\ }\href {https://doi.org/10.1103/PhysRevLett.79.3865} {\bibfield  {journal} {\bibinfo  {journal} {Phys. Rev. Lett.}\ }\textbf {\bibinfo {volume} {79}},\ \bibinfo {pages} {3865} (\bibinfo {year} {1997})}\BibitemShut {NoStop}%
\bibitem [{\citenamefont {Escher}\ \emph {et~al.}(2011)\citenamefont {Escher}, \citenamefont {de~Matos~Filho},\ and\ \citenamefont {Davidovich}}]{Escher2011}%
  \BibitemOpen
  \bibfield  {author} {\bibinfo {author} {\bibfnamefont {B.~M.}\ \bibnamefont {Escher}}, \bibinfo {author} {\bibfnamefont {R.~L.}\ \bibnamefont {de~Matos~Filho}},\ and\ \bibinfo {author} {\bibfnamefont {L.}~\bibnamefont {Davidovich}},\ }\bibfield  {title} {\bibinfo {title} {General framework for estimating the ultimate precision limit in noisy quantum-enhanced metrology},\ }\href {https://doi.org/10.1038/nphys1958} {\bibfield  {journal} {\bibinfo  {journal} {Nat. Phys.}\ }\textbf {\bibinfo {volume} {7}},\ \bibinfo {pages} {406} (\bibinfo {year} {2011})}\BibitemShut {NoStop}%
\bibitem [{\citenamefont {Demkowicz-Dobrzański}\ \emph {et~al.}(2012)\citenamefont {Demkowicz-Dobrzański}, \citenamefont {Kołodyński},\ and\ \citenamefont {Guţă}}]{DemkowiczDobrzanski2012}%
  \BibitemOpen
  \bibfield  {author} {\bibinfo {author} {\bibfnamefont {R.}~\bibnamefont {Demkowicz-Dobrzański}}, \bibinfo {author} {\bibfnamefont {J.}~\bibnamefont {Kołodyński}},\ and\ \bibinfo {author} {\bibfnamefont {M.}~\bibnamefont {Guţă}},\ }\bibfield  {title} {\bibinfo {title} {The elusive heisenberg limit in quantum-enhanced metrology},\ }\href {https://doi.org/10.1038/ncomms2067} {\bibfield  {journal} {\bibinfo  {journal} {Nat. Commun.}\ }\textbf {\bibinfo {volume} {3}},\ \bibinfo {pages} {1063} (\bibinfo {year} {2012})}\BibitemShut {NoStop}%
\bibitem [{\citenamefont {Leibfried}\ \emph {et~al.}(2004)\citenamefont {Leibfried}, \citenamefont {Barrett}, \citenamefont {Schaetz}, \citenamefont {Britton}, \citenamefont {Chiaverini}, \citenamefont {Itano}, \citenamefont {Jost}, \citenamefont {Langer},\ and\ \citenamefont {Wineland}}]{Leibfried2004}%
  \BibitemOpen
  \bibfield  {author} {\bibinfo {author} {\bibfnamefont {D.}~\bibnamefont {Leibfried}}, \bibinfo {author} {\bibfnamefont {M.~D.}\ \bibnamefont {Barrett}}, \bibinfo {author} {\bibfnamefont {T.}~\bibnamefont {Schaetz}}, \bibinfo {author} {\bibfnamefont {J.}~\bibnamefont {Britton}}, \bibinfo {author} {\bibfnamefont {J.}~\bibnamefont {Chiaverini}}, \bibinfo {author} {\bibfnamefont {W.~M.}\ \bibnamefont {Itano}}, \bibinfo {author} {\bibfnamefont {J.~D.}\ \bibnamefont {Jost}}, \bibinfo {author} {\bibfnamefont {C.}~\bibnamefont {Langer}},\ and\ \bibinfo {author} {\bibfnamefont {D.~J.}\ \bibnamefont {Wineland}},\ }\bibfield  {title} {\bibinfo {title} {Toward heisenberg-limited spectroscopy with multiparticle entangled states},\ }\href {https://doi.org/10.1126/science.1097576} {\bibfield  {journal} {\bibinfo  {journal} {Science}\ }\textbf {\bibinfo {volume} {304}},\ \bibinfo {pages} {1476} (\bibinfo {year} {2004})}\BibitemShut {NoStop}%
\bibitem [{\citenamefont {Giovannetti}\ \emph {et~al.}(2004)\citenamefont {Giovannetti}, \citenamefont {Lloyd},\ and\ \citenamefont {Maccone}}]{Giovannetti2004}%
  \BibitemOpen
  \bibfield  {author} {\bibinfo {author} {\bibfnamefont {V.}~\bibnamefont {Giovannetti}}, \bibinfo {author} {\bibfnamefont {S.}~\bibnamefont {Lloyd}},\ and\ \bibinfo {author} {\bibfnamefont {L.}~\bibnamefont {Maccone}},\ }\bibfield  {title} {\bibinfo {title} {Quantum-enhanced measurements: Beating the standard quantum limit},\ }\href {https://doi.org/10.1126/science.1104149} {\bibfield  {journal} {\bibinfo  {journal} {Science}\ }\textbf {\bibinfo {volume} {306}},\ \bibinfo {pages} {1330} (\bibinfo {year} {2004})}\BibitemShut {NoStop}%
\bibitem [{\citenamefont {Giovannetti}\ \emph {et~al.}(2006)\citenamefont {Giovannetti}, \citenamefont {Lloyd},\ and\ \citenamefont {Maccone}}]{Giovannetti2006}%
  \BibitemOpen
  \bibfield  {author} {\bibinfo {author} {\bibfnamefont {V.}~\bibnamefont {Giovannetti}}, \bibinfo {author} {\bibfnamefont {S.}~\bibnamefont {Lloyd}},\ and\ \bibinfo {author} {\bibfnamefont {L.}~\bibnamefont {Maccone}},\ }\bibfield  {title} {\bibinfo {title} {Quantum metrology},\ }\href {https://doi.org/10.1103/PhysRevLett.96.010401} {\bibfield  {journal} {\bibinfo  {journal} {Phys. Rev. Lett.}\ }\textbf {\bibinfo {volume} {96}},\ \bibinfo {pages} {010401} (\bibinfo {year} {2006})}\BibitemShut {NoStop}%
\bibitem [{\citenamefont {Demkowicz-Dobrzanski}\ \emph {et~al.}(2009)\citenamefont {Demkowicz-Dobrzanski}, \citenamefont {Dorner}, \citenamefont {Smith}, \citenamefont {Lundeen}, \citenamefont {Wasilewski}, \citenamefont {Banaszek},\ and\ \citenamefont {Walmsley}}]{DemkowiczDobrzanski2009}%
  \BibitemOpen
  \bibfield  {author} {\bibinfo {author} {\bibfnamefont {R.}~\bibnamefont {Demkowicz-Dobrzanski}}, \bibinfo {author} {\bibfnamefont {U.}~\bibnamefont {Dorner}}, \bibinfo {author} {\bibfnamefont {B.~J.}\ \bibnamefont {Smith}}, \bibinfo {author} {\bibfnamefont {J.~S.}\ \bibnamefont {Lundeen}}, \bibinfo {author} {\bibfnamefont {W.}~\bibnamefont {Wasilewski}}, \bibinfo {author} {\bibfnamefont {K.}~\bibnamefont {Banaszek}},\ and\ \bibinfo {author} {\bibfnamefont {I.~A.}\ \bibnamefont {Walmsley}},\ }\bibfield  {title} {\bibinfo {title} {Quantum phase estimation with lossy interferometers},\ }\href {https://doi.org/10.1103/PhysRevA.80.013825} {\bibfield  {journal} {\bibinfo  {journal} {Phys. Rev. A}\ }\textbf {\bibinfo {volume} {80}},\ \bibinfo {pages} {013825} (\bibinfo {year} {2009})}\BibitemShut {NoStop}%
\bibitem [{\citenamefont {Kacprowicz}\ \emph {et~al.}(2010)\citenamefont {Kacprowicz}, \citenamefont {Demkowicz-Dobrzański}, \citenamefont {Wasilewski}, \citenamefont {Banaszek},\ and\ \citenamefont {Walmsley}}]{Kacprowicz2010}%
  \BibitemOpen
  \bibfield  {author} {\bibinfo {author} {\bibfnamefont {M.}~\bibnamefont {Kacprowicz}}, \bibinfo {author} {\bibfnamefont {R.}~\bibnamefont {Demkowicz-Dobrzański}}, \bibinfo {author} {\bibfnamefont {W.}~\bibnamefont {Wasilewski}}, \bibinfo {author} {\bibfnamefont {K.}~\bibnamefont {Banaszek}},\ and\ \bibinfo {author} {\bibfnamefont {I.~A.}\ \bibnamefont {Walmsley}},\ }\bibfield  {title} {\bibinfo {title} {Experimental quantum-enhanced estimation of a lossy phase shift},\ }\href {https://doi.org/10.1038/nphoton.2010.39} {\bibfield  {journal} {\bibinfo  {journal} {Nat. Photonics}\ }\textbf {\bibinfo {volume} {4}},\ \bibinfo {pages} {357} (\bibinfo {year} {2010})}\BibitemShut {NoStop}%
\bibitem [{\citenamefont {Giovannetti}\ \emph {et~al.}(2011)\citenamefont {Giovannetti}, \citenamefont {Lloyd},\ and\ \citenamefont {Maccone}}]{Giovannetti2011}%
  \BibitemOpen
  \bibfield  {author} {\bibinfo {author} {\bibfnamefont {V.}~\bibnamefont {Giovannetti}}, \bibinfo {author} {\bibfnamefont {S.}~\bibnamefont {Lloyd}},\ and\ \bibinfo {author} {\bibfnamefont {L.}~\bibnamefont {Maccone}},\ }\bibfield  {title} {\bibinfo {title} {Advances in quantum metrology},\ }\href {https://doi.org/10.1038/nphoton.2011.35} {\bibfield  {journal} {\bibinfo  {journal} {Nat. Photonics}\ }\textbf {\bibinfo {volume} {5}},\ \bibinfo {pages} {222} (\bibinfo {year} {2011})}\BibitemShut {NoStop}%
\bibitem [{\citenamefont {Joo}\ \emph {et~al.}(2011)\citenamefont {Joo}, \citenamefont {Munro},\ and\ \citenamefont {Spiller}}]{Joo2011}%
  \BibitemOpen
  \bibfield  {author} {\bibinfo {author} {\bibfnamefont {J.}~\bibnamefont {Joo}}, \bibinfo {author} {\bibfnamefont {W.~J.}\ \bibnamefont {Munro}},\ and\ \bibinfo {author} {\bibfnamefont {T.~P.}\ \bibnamefont {Spiller}},\ }\bibfield  {title} {\bibinfo {title} {Quantum metrology with entangled coherent states},\ }\href {https://doi.org/10.1103/PhysRevLett.107.083601} {\bibfield  {journal} {\bibinfo  {journal} {Phys. Rev. Lett.}\ }\textbf {\bibinfo {volume} {107}},\ \bibinfo {pages} {083601} (\bibinfo {year} {2011})}\BibitemShut {NoStop}%
\bibitem [{\citenamefont {Degen}\ \emph {et~al.}(2017)\citenamefont {Degen}, \citenamefont {Reinhard},\ and\ \citenamefont {Cappellaro}}]{Degen2017}%
  \BibitemOpen
  \bibfield  {author} {\bibinfo {author} {\bibfnamefont {C.~L.}\ \bibnamefont {Degen}}, \bibinfo {author} {\bibfnamefont {F.}~\bibnamefont {Reinhard}},\ and\ \bibinfo {author} {\bibfnamefont {P.}~\bibnamefont {Cappellaro}},\ }\bibfield  {title} {\bibinfo {title} {Quantum sensing},\ }\href {https://doi.org/10.1103/RevModPhys.89.035002} {\bibfield  {journal} {\bibinfo  {journal} {Rev. Mod. Phys.}\ }\textbf {\bibinfo {volume} {89}},\ \bibinfo {pages} {035002} (\bibinfo {year} {2017})}\BibitemShut {NoStop}%
\bibitem [{\citenamefont {Pezz\`e}\ \emph {et~al.}(2018)\citenamefont {Pezz\`e}, \citenamefont {Smerzi}, \citenamefont {Oberthaler}, \citenamefont {Schmied},\ and\ \citenamefont {Treutlein}}]{Pezze2018}%
  \BibitemOpen
  \bibfield  {author} {\bibinfo {author} {\bibfnamefont {L.}~\bibnamefont {Pezz\`e}}, \bibinfo {author} {\bibfnamefont {A.}~\bibnamefont {Smerzi}}, \bibinfo {author} {\bibfnamefont {M.~K.}\ \bibnamefont {Oberthaler}}, \bibinfo {author} {\bibfnamefont {R.}~\bibnamefont {Schmied}},\ and\ \bibinfo {author} {\bibfnamefont {P.}~\bibnamefont {Treutlein}},\ }\bibfield  {title} {\bibinfo {title} {Quantum metrology with nonclassical states of atomic ensembles},\ }\href {https://doi.org/10.1103/RevModPhys.90.035005} {\bibfield  {journal} {\bibinfo  {journal} {Rev. Mod. Phys.}\ }\textbf {\bibinfo {volume} {90}},\ \bibinfo {pages} {035005} (\bibinfo {year} {2018})}\BibitemShut {NoStop}%
\bibitem [{\citenamefont {Daryanoosh}\ \emph {et~al.}(2018)\citenamefont {Daryanoosh}, \citenamefont {Slussarenko}, \citenamefont {Berry}, \citenamefont {Wiseman},\ and\ \citenamefont {Pryde}}]{Daryanoosh2018}%
  \BibitemOpen
  \bibfield  {author} {\bibinfo {author} {\bibfnamefont {S.}~\bibnamefont {Daryanoosh}}, \bibinfo {author} {\bibfnamefont {S.}~\bibnamefont {Slussarenko}}, \bibinfo {author} {\bibfnamefont {D.~W.}\ \bibnamefont {Berry}}, \bibinfo {author} {\bibfnamefont {H.~M.}\ \bibnamefont {Wiseman}},\ and\ \bibinfo {author} {\bibfnamefont {G.~J.}\ \bibnamefont {Pryde}},\ }\bibfield  {title} {\bibinfo {title} {Experimental optical phase measurement approaching the exact heisenberg limit},\ }\href {https://doi.org/10.1038/s41467-018-06601-7} {\bibfield  {journal} {\bibinfo  {journal} {Nat. Commun.}\ }\textbf {\bibinfo {volume} {9}},\ \bibinfo {pages} {4606} (\bibinfo {year} {2018})}\BibitemShut {NoStop}%
\bibitem [{\citenamefont {Barbieri}(2022)}]{Barbieri2022}%
  \BibitemOpen
  \bibfield  {author} {\bibinfo {author} {\bibfnamefont {M.}~\bibnamefont {Barbieri}},\ }\bibfield  {title} {\bibinfo {title} {Optical quantum metrology},\ }\href {https://doi.org/10.1103/PRXQuantum.3.010202} {\bibfield  {journal} {\bibinfo  {journal} {PRX Quantum}\ }\textbf {\bibinfo {volume} {3}},\ \bibinfo {pages} {010202} (\bibinfo {year} {2022})}\BibitemShut {NoStop}%
\bibitem [{\citenamefont {Huang}\ \emph {et~al.}(2024)\citenamefont {Huang}, \citenamefont {Zhuang},\ and\ \citenamefont {Lee}}]{Huang2024}%
  \BibitemOpen
  \bibfield  {author} {\bibinfo {author} {\bibfnamefont {J.}~\bibnamefont {Huang}}, \bibinfo {author} {\bibfnamefont {M.}~\bibnamefont {Zhuang}},\ and\ \bibinfo {author} {\bibfnamefont {C.}~\bibnamefont {Lee}},\ }\bibfield  {title} {\bibinfo {title} {Entanglement-enhanced quantum metrology: From standard quantum limit to heisenberg limit},\ }\href {https://doi.org/10.1063/5.0204102} {\bibfield  {journal} {\bibinfo  {journal} {Appl. Phys. Rev.}\ }\textbf {\bibinfo {volume} {11}},\ \bibinfo {pages} {031302} (\bibinfo {year} {2024})}\BibitemShut {NoStop}%
\bibitem [{\citenamefont {D\"ur}\ \emph {et~al.}(2002)\citenamefont {D\"ur}, \citenamefont {Simon},\ and\ \citenamefont {Cirac}}]{Duer2002}%
  \BibitemOpen
  \bibfield  {author} {\bibinfo {author} {\bibfnamefont {W.}~\bibnamefont {D\"ur}}, \bibinfo {author} {\bibfnamefont {C.}~\bibnamefont {Simon}},\ and\ \bibinfo {author} {\bibfnamefont {J.~I.}\ \bibnamefont {Cirac}},\ }\bibfield  {title} {\bibinfo {title} {Effective size of certain macroscopic quantum superpositions},\ }\href {https://doi.org/10.1103/PhysRevLett.89.210402} {\bibfield  {journal} {\bibinfo  {journal} {Phys. Rev. Lett.}\ }\textbf {\bibinfo {volume} {89}},\ \bibinfo {pages} {210402} (\bibinfo {year} {2002})}\BibitemShut {NoStop}%
\bibitem [{\citenamefont {D\"ur}\ and\ \citenamefont {Briegel}(2004)}]{Duer2004}%
  \BibitemOpen
  \bibfield  {author} {\bibinfo {author} {\bibfnamefont {W.}~\bibnamefont {D\"ur}}\ and\ \bibinfo {author} {\bibfnamefont {H.-J.}\ \bibnamefont {Briegel}},\ }\bibfield  {title} {\bibinfo {title} {Stability of macroscopic entanglement under decoherence},\ }\href {https://doi.org/10.1103/PhysRevLett.92.180403} {\bibfield  {journal} {\bibinfo  {journal} {Phys. Rev. Lett.}\ }\textbf {\bibinfo {volume} {92}},\ \bibinfo {pages} {180403} (\bibinfo {year} {2004})}\BibitemShut {NoStop}%
\bibitem [{\citenamefont {Aolita}\ \emph {et~al.}(2008)\citenamefont {Aolita}, \citenamefont {Chaves}, \citenamefont {Cavalcanti}, \citenamefont {Ac\'{\i}n},\ and\ \citenamefont {Davidovich}}]{Aolita2008}%
  \BibitemOpen
  \bibfield  {author} {\bibinfo {author} {\bibfnamefont {L.}~\bibnamefont {Aolita}}, \bibinfo {author} {\bibfnamefont {R.}~\bibnamefont {Chaves}}, \bibinfo {author} {\bibfnamefont {D.}~\bibnamefont {Cavalcanti}}, \bibinfo {author} {\bibfnamefont {A.}~\bibnamefont {Ac\'{\i}n}},\ and\ \bibinfo {author} {\bibfnamefont {L.}~\bibnamefont {Davidovich}},\ }\bibfield  {title} {\bibinfo {title} {Scaling laws for the decay of multiqubit entanglement},\ }\href {https://doi.org/10.1103/PhysRevLett.100.080501} {\bibfield  {journal} {\bibinfo  {journal} {Phys. Rev. Lett.}\ }\textbf {\bibinfo {volume} {100}},\ \bibinfo {pages} {080501} (\bibinfo {year} {2008})}\BibitemShut {NoStop}%
\bibitem [{\citenamefont {Lu}\ \emph {et~al.}(2014)\citenamefont {Lu}, \citenamefont {Chen}, \citenamefont {Liu}, \citenamefont {Xu}, \citenamefont {Yao}, \citenamefont {Li}, \citenamefont {Liu}, \citenamefont {Zhao}, \citenamefont {Chen},\ and\ \citenamefont {Pan}}]{Lu2014}%
  \BibitemOpen
  \bibfield  {author} {\bibinfo {author} {\bibfnamefont {H.}~\bibnamefont {Lu}}, \bibinfo {author} {\bibfnamefont {L.-K.}\ \bibnamefont {Chen}}, \bibinfo {author} {\bibfnamefont {C.}~\bibnamefont {Liu}}, \bibinfo {author} {\bibfnamefont {P.}~\bibnamefont {Xu}}, \bibinfo {author} {\bibfnamefont {X.-C.}\ \bibnamefont {Yao}}, \bibinfo {author} {\bibfnamefont {L.}~\bibnamefont {Li}}, \bibinfo {author} {\bibfnamefont {N.-L.}\ \bibnamefont {Liu}}, \bibinfo {author} {\bibfnamefont {B.}~\bibnamefont {Zhao}}, \bibinfo {author} {\bibfnamefont {Y.-A.}\ \bibnamefont {Chen}},\ and\ \bibinfo {author} {\bibfnamefont {J.-W.}\ \bibnamefont {Pan}},\ }\bibfield  {title} {\bibinfo {title} {Experimental realization of a concatenated greenberger-horne-zeilinger state for macroscopic quantum superpositions},\ }\href {https://doi.org/10.1038/nphoton.2014.81} {\bibfield  {journal} {\bibinfo  {journal} {Nat. Photonics}\ }\textbf {\bibinfo {volume} {8}},\ \bibinfo {pages} {364} (\bibinfo {year} {2014})}\BibitemShut {NoStop}%
\bibitem [{\citenamefont {Wang}\ \emph {et~al.}(2024)\citenamefont {Wang}, \citenamefont {Yang}, \citenamefont {Ji}, \citenamefont {Liu}, \citenamefont {Dong},\ and\ \citenamefont {Xiu}}]{Wang2024}%
  \BibitemOpen
  \bibfield  {author} {\bibinfo {author} {\bibfnamefont {J.~P.}\ \bibnamefont {Wang}}, \bibinfo {author} {\bibfnamefont {L.~P.}\ \bibnamefont {Yang}}, \bibinfo {author} {\bibfnamefont {Y.~Q.}\ \bibnamefont {Ji}}, \bibinfo {author} {\bibfnamefont {Y.~L.}\ \bibnamefont {Liu}}, \bibinfo {author} {\bibfnamefont {L.}~\bibnamefont {Dong}},\ and\ \bibinfo {author} {\bibfnamefont {X.~M.}\ \bibnamefont {Xiu}},\ }\bibfield  {title} {\bibinfo {title} {Fast generation of ghz state by designing the evolution operators with rydberg superatom},\ }\href {https://doi.org/10.1007/s11128-024-04587-4} {\bibfield  {journal} {\bibinfo  {journal} {Quantum Inf. Process.}\ }\textbf {\bibinfo {volume} {23}},\ \bibinfo {pages} {377} (\bibinfo {year} {2024})}\BibitemShut {NoStop}%
\bibitem [{\citenamefont {Davis}\ \emph {et~al.}(2017)\citenamefont {Davis}, \citenamefont {Bentsen}, \citenamefont {Li},\ and\ \citenamefont {Schleier-Smith}}]{Davis2017}%
  \BibitemOpen
  \bibfield  {author} {\bibinfo {author} {\bibfnamefont {E.}~\bibnamefont {Davis}}, \bibinfo {author} {\bibfnamefont {G.}~\bibnamefont {Bentsen}}, \bibinfo {author} {\bibfnamefont {T.}~\bibnamefont {Li}},\ and\ \bibinfo {author} {\bibfnamefont {M.}~\bibnamefont {Schleier-Smith}},\ }\bibfield  {title} {\bibinfo {title} {Advantages of interaction-based readout for quantum sensing}\ }(\bibinfo {year} {2017})\ p.\ \bibinfo {pages} {101180Z}\BibitemShut {NoStop}%
\bibitem [{\citenamefont {Nolan}\ \emph {et~al.}(2017)\citenamefont {Nolan}, \citenamefont {Szigeti},\ and\ \citenamefont {Haine}}]{Nolan2017}%
  \BibitemOpen
  \bibfield  {author} {\bibinfo {author} {\bibfnamefont {S.~P.}\ \bibnamefont {Nolan}}, \bibinfo {author} {\bibfnamefont {S.~S.}\ \bibnamefont {Szigeti}},\ and\ \bibinfo {author} {\bibfnamefont {S.~A.}\ \bibnamefont {Haine}},\ }\bibfield  {title} {\bibinfo {title} {Optimal and robust quantum metrology using interaction-based readouts},\ }\href {https://doi.org/10.1103/PhysRevLett.119.193601} {\bibfield  {journal} {\bibinfo  {journal} {Phys. Rev. Lett.}\ }\textbf {\bibinfo {volume} {119}},\ \bibinfo {pages} {193601} (\bibinfo {year} {2017})}\BibitemShut {NoStop}%
\bibitem [{\citenamefont {Braunstein}\ and\ \citenamefont {Caves}(1994)}]{Braunstein1994}%
  \BibitemOpen
  \bibfield  {author} {\bibinfo {author} {\bibfnamefont {S.~L.}\ \bibnamefont {Braunstein}}\ and\ \bibinfo {author} {\bibfnamefont {C.~M.}\ \bibnamefont {Caves}},\ }\bibfield  {title} {\bibinfo {title} {Statistical distance and the geometry of quantum states},\ }\href {https://doi.org/10.1103/PhysRevLett.72.3439} {\bibfield  {journal} {\bibinfo  {journal} {Phys. Rev. Lett.}\ }\textbf {\bibinfo {volume} {72}},\ \bibinfo {pages} {3439} (\bibinfo {year} {1994})}\BibitemShut {NoStop}%
\bibitem [{\citenamefont {Maccone}\ and\ \citenamefont {Giovannetti}(2011)}]{Maccone2011}%
  \BibitemOpen
  \bibfield  {author} {\bibinfo {author} {\bibfnamefont {L.}~\bibnamefont {Maccone}}\ and\ \bibinfo {author} {\bibfnamefont {V.}~\bibnamefont {Giovannetti}},\ }\bibfield  {title} {\bibinfo {title} {Beauty and the noisy beast},\ }\href {https://doi.org/10.1038/nphys1976} {\bibfield  {journal} {\bibinfo  {journal} {Nat. Phys.}\ }\textbf {\bibinfo {volume} {7}},\ \bibinfo {pages} {376} (\bibinfo {year} {2011})}\BibitemShut {NoStop}%
\bibitem [{\citenamefont {Smirne}\ \emph {et~al.}(2016)\citenamefont {Smirne}, \citenamefont {Ko\l{}ody\ifmmode~\acute{n}\else \'{n}\fi{}ski}, \citenamefont {Huelga},\ and\ \citenamefont {Demkowicz-Dobrza\ifmmode~\acute{n}\else \'{n}\fi{}ski}}]{Smirne2016}%
  \BibitemOpen
  \bibfield  {author} {\bibinfo {author} {\bibfnamefont {A.}~\bibnamefont {Smirne}}, \bibinfo {author} {\bibfnamefont {J.}~\bibnamefont {Ko\l{}ody\ifmmode~\acute{n}\else \'{n}\fi{}ski}}, \bibinfo {author} {\bibfnamefont {S.~F.}\ \bibnamefont {Huelga}},\ and\ \bibinfo {author} {\bibfnamefont {R.}~\bibnamefont {Demkowicz-Dobrza\ifmmode~\acute{n}\else \'{n}\fi{}ski}},\ }\bibfield  {title} {\bibinfo {title} {Ultimate precision limits for noisy frequency estimation},\ }\href {https://doi.org/10.1103/PhysRevLett.116.120801} {\bibfield  {journal} {\bibinfo  {journal} {Phys. Rev. Lett.}\ }\textbf {\bibinfo {volume} {116}},\ \bibinfo {pages} {120801} (\bibinfo {year} {2016})}\BibitemShut {NoStop}%
\bibitem [{\citenamefont {Thomas-Peter}\ \emph {et~al.}(2011)\citenamefont {Thomas-Peter}, \citenamefont {Smith}, \citenamefont {Datta}, \citenamefont {Zhang}, \citenamefont {Dorner},\ and\ \citenamefont {Walmsley}}]{ThomasPeter2011}%
  \BibitemOpen
  \bibfield  {author} {\bibinfo {author} {\bibfnamefont {N.}~\bibnamefont {Thomas-Peter}}, \bibinfo {author} {\bibfnamefont {B.~J.}\ \bibnamefont {Smith}}, \bibinfo {author} {\bibfnamefont {A.}~\bibnamefont {Datta}}, \bibinfo {author} {\bibfnamefont {L.}~\bibnamefont {Zhang}}, \bibinfo {author} {\bibfnamefont {U.}~\bibnamefont {Dorner}},\ and\ \bibinfo {author} {\bibfnamefont {I.~A.}\ \bibnamefont {Walmsley}},\ }\bibfield  {title} {\bibinfo {title} {Real-world quantum sensors: Evaluating resources for precision measurement},\ }\href {https://doi.org/10.1103/PhysRevLett.107.113603} {\bibfield  {journal} {\bibinfo  {journal} {Phys. Rev. Lett.}\ }\textbf {\bibinfo {volume} {107}},\ \bibinfo {pages} {113603} (\bibinfo {year} {2011})}\BibitemShut {NoStop}%
\bibitem [{\citenamefont {Albarelli}\ \emph {et~al.}(2018)\citenamefont {Albarelli}, \citenamefont {Rossi}, \citenamefont {Tamascelli},\ and\ \citenamefont {Genoni}}]{Albarelli2018}%
  \BibitemOpen
  \bibfield  {author} {\bibinfo {author} {\bibfnamefont {F.}~\bibnamefont {Albarelli}}, \bibinfo {author} {\bibfnamefont {M.~A.~C.}\ \bibnamefont {Rossi}}, \bibinfo {author} {\bibfnamefont {D.}~\bibnamefont {Tamascelli}},\ and\ \bibinfo {author} {\bibfnamefont {M.~G.}\ \bibnamefont {Genoni}},\ }\bibfield  {title} {\bibinfo {title} {Restoring {H}eisenberg scaling in noisy quantum metrology by monitoring the environment},\ }\href {https://doi.org/10.22331/q-2018-12-03-110} {\bibfield  {journal} {\bibinfo  {journal} {Quantum}\ }\textbf {\bibinfo {volume} {2}},\ \bibinfo {pages} {110} (\bibinfo {year} {2018})}\BibitemShut {NoStop}%
\bibitem [{\citenamefont {Albarelli}\ and\ \citenamefont {Demkowicz-Dobrza\ifmmode~\acute{n}\else \'{n}\fi{}ski}(2022)}]{Albarelli2022}%
  \BibitemOpen
  \bibfield  {author} {\bibinfo {author} {\bibfnamefont {F.}~\bibnamefont {Albarelli}}\ and\ \bibinfo {author} {\bibfnamefont {R.}~\bibnamefont {Demkowicz-Dobrza\ifmmode~\acute{n}\else \'{n}\fi{}ski}},\ }\bibfield  {title} {\bibinfo {title} {Probe incompatibility in multiparameter noisy quantum metrology},\ }\href {https://doi.org/10.1103/PhysRevX.12.011039} {\bibfield  {journal} {\bibinfo  {journal} {Phys. Rev. X}\ }\textbf {\bibinfo {volume} {12}},\ \bibinfo {pages} {011039} (\bibinfo {year} {2022})}\BibitemShut {NoStop}%
\bibitem [{\citenamefont {G\'orecki}\ \emph {et~al.}(2022)\citenamefont {G\'orecki}, \citenamefont {Riccardi},\ and\ \citenamefont {Maccone}}]{Gorecki2022}%
  \BibitemOpen
  \bibfield  {author} {\bibinfo {author} {\bibfnamefont {W.}~\bibnamefont {G\'orecki}}, \bibinfo {author} {\bibfnamefont {A.}~\bibnamefont {Riccardi}},\ and\ \bibinfo {author} {\bibfnamefont {L.}~\bibnamefont {Maccone}},\ }\bibfield  {title} {\bibinfo {title} {Quantum metrology of noisy spreading channels},\ }\href {https://doi.org/10.1103/PhysRevLett.129.240503} {\bibfield  {journal} {\bibinfo  {journal} {Phys. Rev. Lett.}\ }\textbf {\bibinfo {volume} {129}},\ \bibinfo {pages} {240503} (\bibinfo {year} {2022})}\BibitemShut {NoStop}%
\bibitem [{\citenamefont {Bai}\ and\ \citenamefont {An}(2023)}]{Bai2023}%
  \BibitemOpen
  \bibfield  {author} {\bibinfo {author} {\bibfnamefont {S.-Y.}\ \bibnamefont {Bai}}\ and\ \bibinfo {author} {\bibfnamefont {J.-H.}\ \bibnamefont {An}},\ }\bibfield  {title} {\bibinfo {title} {Floquet engineering to overcome no-go theorem of noisy quantum metrology},\ }\href {https://doi.org/10.1103/PhysRevLett.131.050801} {\bibfield  {journal} {\bibinfo  {journal} {Phys. Rev. Lett.}\ }\textbf {\bibinfo {volume} {131}},\ \bibinfo {pages} {050801} (\bibinfo {year} {2023})}\BibitemShut {NoStop}%
\bibitem [{\citenamefont {Zhou}(2024)}]{Zhou2024}%
  \BibitemOpen
  \bibfield  {author} {\bibinfo {author} {\bibfnamefont {S.}~\bibnamefont {Zhou}},\ }\bibfield  {title} {\bibinfo {title} {Limits of noisy quantum metrology with restricted quantum controls},\ }\href {https://doi.org/10.1103/PhysRevLett.133.170801} {\bibfield  {journal} {\bibinfo  {journal} {Phys. Rev. Lett.}\ }\textbf {\bibinfo {volume} {133}},\ \bibinfo {pages} {170801} (\bibinfo {year} {2024})}\BibitemShut {NoStop}%
\bibitem [{\citenamefont {Zou}\ and\ \citenamefont {Wang}(2022)}]{Zou2022}%
  \BibitemOpen
  \bibfield  {author} {\bibinfo {author} {\bibfnamefont {Z.}~\bibnamefont {Zou}}\ and\ \bibinfo {author} {\bibfnamefont {J.}~\bibnamefont {Wang}},\ }\bibfield  {title} {\bibinfo {title} {Pseudoclassical dynamics of the kicked top},\ }\bibfield  {journal} {\bibinfo  {journal} {Entropy}\ }\textbf {\bibinfo {volume} {24}},\ \href {https://doi.org/10.3390/e24081092} {10.3390/e24081092} (\bibinfo {year} {2022})\BibitemShut {NoStop}%
\bibitem [{\citenamefont {PARIS}(2009)}]{PARIS2009}%
  \BibitemOpen
  \bibfield  {author} {\bibinfo {author} {\bibfnamefont {M.~A. T. T. E. O. G.~A.}\ \bibnamefont {PARIS}},\ }\bibfield  {title} {\bibinfo {title} {Quantum estimation for quantum technology},\ }\href {https://doi.org/10.1142/S0219749909004839} {\bibfield  {journal} {\bibinfo  {journal} {Int. J. Quantum Inform.}\ }\textbf {\bibinfo {volume} {07}},\ \bibinfo {pages} {125} (\bibinfo {year} {2009})}\BibitemShut {NoStop}%
\bibitem [{\citenamefont {Liu}\ \emph {et~al.}(2016)\citenamefont {Liu}, \citenamefont {Chen}, \citenamefont {Jing},\ and\ \citenamefont {Wang}}]{Liu2016}%
  \BibitemOpen
  \bibfield  {author} {\bibinfo {author} {\bibfnamefont {J.}~\bibnamefont {Liu}}, \bibinfo {author} {\bibfnamefont {J.}~\bibnamefont {Chen}}, \bibinfo {author} {\bibfnamefont {X.-X.}\ \bibnamefont {Jing}},\ and\ \bibinfo {author} {\bibfnamefont {X.}~\bibnamefont {Wang}},\ }\bibfield  {title} {\bibinfo {title} {Quantum fisher information and symmetric logarithmic derivative via anti-commutators},\ }\href {https://doi.org/10.1088/1751-8113/49/27/275302} {\bibfield  {journal} {\bibinfo  {journal} {J. Phys. A: Math. Theor.}\ }\textbf {\bibinfo {volume} {49}},\ \bibinfo {pages} {275302} (\bibinfo {year} {2016})}\BibitemShut {NoStop}%
\bibitem [{\citenamefont {Gorin}\ \emph {et~al.}(2006)\citenamefont {Gorin}, \citenamefont {Prosen}, \citenamefont {Seligman},\ and\ \citenamefont {Žnidarič}}]{Gorin2006}%
  \BibitemOpen
  \bibfield  {author} {\bibinfo {author} {\bibfnamefont {T.}~\bibnamefont {Gorin}}, \bibinfo {author} {\bibfnamefont {T.}~\bibnamefont {Prosen}}, \bibinfo {author} {\bibfnamefont {T.~H.}\ \bibnamefont {Seligman}},\ and\ \bibinfo {author} {\bibfnamefont {M.}~\bibnamefont {Žnidarič}},\ }\bibfield  {title} {\bibinfo {title} {Dynamics of loschmidt echoes and fidelity decay},\ }\href {https://doi.org/10.1016/j.physrep.2006.09.003} {\bibfield  {journal} {\bibinfo  {journal} {Phys. Rep.}\ }\textbf {\bibinfo {volume} {435}},\ \bibinfo {pages} {33} (\bibinfo {year} {2006})}\BibitemShut {NoStop}%
\bibitem [{\citenamefont {Fiderer}\ and\ \citenamefont {Braun}(2018)}]{Fiderer2018}%
  \BibitemOpen
  \bibfield  {author} {\bibinfo {author} {\bibfnamefont {L.~J.}\ \bibnamefont {Fiderer}}\ and\ \bibinfo {author} {\bibfnamefont {D.}~\bibnamefont {Braun}},\ }\bibfield  {title} {\bibinfo {title} {Quantum metrology with quantum-chaotic sensors},\ }\href {https://doi.org/10.1038/s41467-018-03623-z} {\bibfield  {journal} {\bibinfo  {journal} {Nat. Commun.}\ }\textbf {\bibinfo {volume} {9}},\ \bibinfo {pages} {1351} (\bibinfo {year} {2018})}\BibitemShut {NoStop}%
\bibitem [{\citenamefont {Miszczak}\ \emph {et~al.}(2009)\citenamefont {Miszczak}, \citenamefont {Pucha\l{}a}, \citenamefont {Horodecki}, \citenamefont {Uhlmann},\ and\ \citenamefont {Zyczkowski}}]{Miszczak2009}%
  \BibitemOpen
  \bibfield  {author} {\bibinfo {author} {\bibfnamefont {J.~A.}\ \bibnamefont {Miszczak}}, \bibinfo {author} {\bibfnamefont {Z.}~\bibnamefont {Pucha\l{}a}}, \bibinfo {author} {\bibfnamefont {P.}~\bibnamefont {Horodecki}}, \bibinfo {author} {\bibfnamefont {A.}~\bibnamefont {Uhlmann}},\ and\ \bibinfo {author} {\bibfnamefont {K.}~\bibnamefont {Zyczkowski}},\ }\bibfield  {title} {\bibinfo {title} {Sub- and super-fidelity as bounds for quantum fidelity},\ }\href@noop {} {\bibfield  {journal} {\bibinfo  {journal} {Quantum Info. Comput.}\ }\textbf {\bibinfo {volume} {9}},\ \bibinfo {pages} {103–130} (\bibinfo {year} {2009})}\BibitemShut {NoStop}%
\bibitem [{\citenamefont {Lerose}\ and\ \citenamefont {Pappalardi}(2020)}]{Lerose2020}%
  \BibitemOpen
  \bibfield  {author} {\bibinfo {author} {\bibfnamefont {A.}~\bibnamefont {Lerose}}\ and\ \bibinfo {author} {\bibfnamefont {S.}~\bibnamefont {Pappalardi}},\ }\bibfield  {title} {\bibinfo {title} {Bridging entanglement dynamics and chaos in semiclassical systems},\ }\href {https://doi.org/10.1103/PhysRevA.102.032404} {\bibfield  {journal} {\bibinfo  {journal} {Phys. Rev. A}\ }\textbf {\bibinfo {volume} {102}},\ \bibinfo {pages} {032404} (\bibinfo {year} {2020})}\BibitemShut {NoStop}%
\bibitem [{\citenamefont {Liu}\ \emph {et~al.}(2021)\citenamefont {Liu}, \citenamefont {Zhuang}, \citenamefont {Zhu}, \citenamefont {Huang},\ and\ \citenamefont {Lee}}]{Liu2021}%
  \BibitemOpen
  \bibfield  {author} {\bibinfo {author} {\bibfnamefont {W.}~\bibnamefont {Liu}}, \bibinfo {author} {\bibfnamefont {M.}~\bibnamefont {Zhuang}}, \bibinfo {author} {\bibfnamefont {B.}~\bibnamefont {Zhu}}, \bibinfo {author} {\bibfnamefont {J.}~\bibnamefont {Huang}},\ and\ \bibinfo {author} {\bibfnamefont {C.}~\bibnamefont {Lee}},\ }\bibfield  {title} {\bibinfo {title} {Quantum metrology via chaos in a driven bose-josephson system},\ }\href {https://doi.org/10.1103/PhysRevA.103.023309} {\bibfield  {journal} {\bibinfo  {journal} {Phys. Rev. A}\ }\textbf {\bibinfo {volume} {103}},\ \bibinfo {pages} {023309} (\bibinfo {year} {2021})}\BibitemShut {NoStop}%
\bibitem [{\citenamefont {Zaslavsky}(1981)}]{Zaslavsky1981}%
  \BibitemOpen
  \bibfield  {author} {\bibinfo {author} {\bibfnamefont {G.~M.}\ \bibnamefont {Zaslavsky}},\ }\bibfield  {title} {\bibinfo {title} {Stochasticity in quantum systems},\ }\href {https://doi.org/10.1016/0370-1573(81)90127-7} {\bibfield  {journal} {\bibinfo  {journal} {Physics Reports}\ }\textbf {\bibinfo {volume} {80}},\ \bibinfo {pages} {157} (\bibinfo {year} {1981})}\BibitemShut {NoStop}%
\bibitem [{\citenamefont {Silvestrov}\ and\ \citenamefont {Beenakker}(2002)}]{Silvestrov2002}%
  \BibitemOpen
  \bibfield  {author} {\bibinfo {author} {\bibfnamefont {P.~G.}\ \bibnamefont {Silvestrov}}\ and\ \bibinfo {author} {\bibfnamefont {C.~W.~J.}\ \bibnamefont {Beenakker}},\ }\bibfield  {title} {\bibinfo {title} {Ehrenfest times for classically chaotic systems},\ }\href {https://doi.org/10.1103/PhysRevE.65.035208} {\bibfield  {journal} {\bibinfo  {journal} {Phys. Rev. E}\ }\textbf {\bibinfo {volume} {65}},\ \bibinfo {pages} {035208} (\bibinfo {year} {2002})}\BibitemShut {NoStop}%
\bibitem [{\citenamefont {Haake}\ \emph {et~al.}(1987)\citenamefont {Haake}, \citenamefont {Kuś},\ and\ \citenamefont {Scharf}}]{Haake1987}%
  \BibitemOpen
  \bibfield  {author} {\bibinfo {author} {\bibfnamefont {F.}~\bibnamefont {Haake}}, \bibinfo {author} {\bibfnamefont {M.}~\bibnamefont {Kuś}},\ and\ \bibinfo {author} {\bibfnamefont {R.}~\bibnamefont {Scharf}},\ }\bibfield  {title} {\bibinfo {title} {Classical and quantum chaos for a kicked top},\ }\href {https://doi.org/10.1007/BF01303727} {\bibfield  {journal} {\bibinfo  {journal} {Z. Phys. B Condens. Matter}\ }\textbf {\bibinfo {volume} {65}},\ \bibinfo {pages} {381} (\bibinfo {year} {1987})}\BibitemShut {NoStop}%
\bibitem [{\citenamefont {Chaudhury}\ \emph {et~al.}(2009)\citenamefont {Chaudhury}, \citenamefont {Smith}, \citenamefont {Anderson}, \citenamefont {Ghose},\ and\ \citenamefont {Jessen}}]{Chaudhury2009}%
  \BibitemOpen
  \bibfield  {author} {\bibinfo {author} {\bibfnamefont {S.}~\bibnamefont {Chaudhury}}, \bibinfo {author} {\bibfnamefont {A.}~\bibnamefont {Smith}}, \bibinfo {author} {\bibfnamefont {B.~E.}\ \bibnamefont {Anderson}}, \bibinfo {author} {\bibfnamefont {S.}~\bibnamefont {Ghose}},\ and\ \bibinfo {author} {\bibfnamefont {P.~S.}\ \bibnamefont {Jessen}},\ }\bibfield  {title} {\bibinfo {title} {Quantum signatures of chaos in a kicked top},\ }\href {https://doi.org/10.1038/nature08396} {\bibfield  {journal} {\bibinfo  {journal} {Nature}\ }\textbf {\bibinfo {volume} {461}},\ \bibinfo {pages} {768} (\bibinfo {year} {2009})}\BibitemShut {NoStop}%
\bibitem [{\citenamefont {Xu}\ \emph {et~al.}(2010)\citenamefont {Xu}, \citenamefont {Zhou}, \citenamefont {Xie},\ and\ \citenamefont {Feng}}]{Xu2010}%
  \BibitemOpen
  \bibfield  {author} {\bibinfo {author} {\bibfnamefont {Y.~Y.}\ \bibnamefont {Xu}}, \bibinfo {author} {\bibfnamefont {F.}~\bibnamefont {Zhou}}, \bibinfo {author} {\bibfnamefont {Y.}~\bibnamefont {Xie}},\ and\ \bibinfo {author} {\bibfnamefont {M.}~\bibnamefont {Feng}},\ }\bibfield  {title} {\bibinfo {title} {Exploring the quantum kicked top by ion-trap quantum computing},\ }\href {https://doi.org/10.1088/0953-4075/43/18/185503} {\bibfield  {journal} {\bibinfo  {journal} {J. Phys. B: At. Mol. Opt. Phys.}\ }\textbf {\bibinfo {volume} {43}},\ \bibinfo {pages} {185503} (\bibinfo {year} {2010})}\BibitemShut {NoStop}%
\bibitem [{\citenamefont {Sieberer}\ \emph {et~al.}(2019)\citenamefont {Sieberer}, \citenamefont {Olsacher}, \citenamefont {Elben}, \citenamefont {Heyl}, \citenamefont {Hauke}, \citenamefont {Haake},\ and\ \citenamefont {Zoller}}]{Sieberer2019}%
  \BibitemOpen
  \bibfield  {author} {\bibinfo {author} {\bibfnamefont {L.~M.}\ \bibnamefont {Sieberer}}, \bibinfo {author} {\bibfnamefont {T.}~\bibnamefont {Olsacher}}, \bibinfo {author} {\bibfnamefont {A.}~\bibnamefont {Elben}}, \bibinfo {author} {\bibfnamefont {M.}~\bibnamefont {Heyl}}, \bibinfo {author} {\bibfnamefont {P.}~\bibnamefont {Hauke}}, \bibinfo {author} {\bibfnamefont {F.}~\bibnamefont {Haake}},\ and\ \bibinfo {author} {\bibfnamefont {P.}~\bibnamefont {Zoller}},\ }\bibfield  {title} {\bibinfo {title} {Digital quantum simulation, trotter errors, and quantum chaos of the kicked top},\ }\href {https://doi.org/10.1038/s41534-019-0192-5} {\bibfield  {journal} {\bibinfo  {journal} {npj Quantum Information}\ }\textbf {\bibinfo {volume} {5}},\ \bibinfo {pages} {78} (\bibinfo {year} {2019})}\BibitemShut {NoStop}%
\bibitem [{\citenamefont {Izrailev}\ and\ \citenamefont {Shepelyanskii}(1980)}]{Izrailev1980}%
  \BibitemOpen
  \bibfield  {author} {\bibinfo {author} {\bibfnamefont {F.~M.}\ \bibnamefont {Izrailev}}\ and\ \bibinfo {author} {\bibfnamefont {D.~L.}\ \bibnamefont {Shepelyanskii}},\ }\bibfield  {title} {\bibinfo {title} {Quantum resonance for a rotator in a nonlinear periodic field},\ }\href {https://doi.org/10.1007/BF01029131} {\bibfield  {journal} {\bibinfo  {journal} {Theor. Math. Phys.}\ }\textbf {\bibinfo {volume} {43}},\ \bibinfo {pages} {553} (\bibinfo {year} {1980})}\BibitemShut {NoStop}%
\bibitem [{\citenamefont {Wimberger}\ \emph {et~al.}(2003)\citenamefont {Wimberger}, \citenamefont {Guarneri},\ and\ \citenamefont {Fishman}}]{Wimberger2003}%
  \BibitemOpen
  \bibfield  {author} {\bibinfo {author} {\bibfnamefont {S.}~\bibnamefont {Wimberger}}, \bibinfo {author} {\bibfnamefont {I.}~\bibnamefont {Guarneri}},\ and\ \bibinfo {author} {\bibfnamefont {S.}~\bibnamefont {Fishman}},\ }\bibfield  {title} {\bibinfo {title} {Quantum resonances and decoherence for $\delta$-kicked atoms},\ }\href {https://doi.org/10.1088/0951-7715/16/4/312} {\bibfield  {journal} {\bibinfo  {journal} {Nonlinearity}\ }\textbf {\bibinfo {volume} {16}},\ \bibinfo {pages} {1381} (\bibinfo {year} {2003})}\BibitemShut {NoStop}%
\bibitem [{\citenamefont {Wimberger}\ \emph {et~al.}(2004)\citenamefont {Wimberger}, \citenamefont {Guarneri},\ and\ \citenamefont {Fishman}}]{Wimberger2004}%
  \BibitemOpen
  \bibfield  {author} {\bibinfo {author} {\bibfnamefont {S.}~\bibnamefont {Wimberger}}, \bibinfo {author} {\bibfnamefont {I.}~\bibnamefont {Guarneri}},\ and\ \bibinfo {author} {\bibfnamefont {S.}~\bibnamefont {Fishman}},\ }\bibfield  {title} {\bibinfo {title} {Classical scaling theory of quantum resonances},\ }\href {https://doi.org/10.1103/PhysRevLett.92.084102} {\bibfield  {journal} {\bibinfo  {journal} {Phys. Rev. Lett.}\ }\textbf {\bibinfo {volume} {92}},\ \bibinfo {pages} {084102} (\bibinfo {year} {2004})}\BibitemShut {NoStop}%
\bibitem [{\citenamefont {Wimberger}\ and\ \citenamefont {Sadgrove}(2005)}]{Wimberger2005}%
  \BibitemOpen
  \bibfield  {author} {\bibinfo {author} {\bibfnamefont {S.}~\bibnamefont {Wimberger}}\ and\ \bibinfo {author} {\bibfnamefont {M.}~\bibnamefont {Sadgrove}},\ }\bibfield  {title} {\bibinfo {title} {The role of quasi-momentum in the resonant dynamics of the atom-optics kicked rotor},\ }\href {https://doi.org/10.1088/0305-4470/38/49/007} {\bibfield  {journal} {\bibinfo  {journal} {J. Phys. A: Math. Gen.}\ }\textbf {\bibinfo {volume} {38}},\ \bibinfo {pages} {10549} (\bibinfo {year} {2005})}\BibitemShut {NoStop}%
\bibitem [{\citenamefont {Dana}\ and\ \citenamefont {Dorofeev}(2006)}]{Dana2006}%
  \BibitemOpen
  \bibfield  {author} {\bibinfo {author} {\bibfnamefont {I.}~\bibnamefont {Dana}}\ and\ \bibinfo {author} {\bibfnamefont {D.~L.}\ \bibnamefont {Dorofeev}},\ }\bibfield  {title} {\bibinfo {title} {General quantum resonances of the kicked particle},\ }\href {https://doi.org/10.1103/PhysRevE.73.026206} {\bibfield  {journal} {\bibinfo  {journal} {Phys. Rev. E}\ }\textbf {\bibinfo {volume} {73}},\ \bibinfo {pages} {026206} (\bibinfo {year} {2006})}\BibitemShut {NoStop}%
\bibitem [{\citenamefont {Abb}\ \emph {et~al.}(2009)\citenamefont {Abb}, \citenamefont {Guarneri},\ and\ \citenamefont {Wimberger}}]{Abb2009}%
  \BibitemOpen
  \bibfield  {author} {\bibinfo {author} {\bibfnamefont {M.}~\bibnamefont {Abb}}, \bibinfo {author} {\bibfnamefont {I.}~\bibnamefont {Guarneri}},\ and\ \bibinfo {author} {\bibfnamefont {S.}~\bibnamefont {Wimberger}},\ }\bibfield  {title} {\bibinfo {title} {Pseudoclassical theory for fidelity of nearly resonant quantum rotors},\ }\href {https://doi.org/10.1103/PhysRevE.80.035206} {\bibfield  {journal} {\bibinfo  {journal} {Phys. Rev. E}\ }\textbf {\bibinfo {volume} {80}},\ \bibinfo {pages} {035206} (\bibinfo {year} {2009})}\BibitemShut {NoStop}%
\bibitem [{\citenamefont {McDowall}\ \emph {et~al.}(2009)\citenamefont {McDowall}, \citenamefont {Hilliard}, \citenamefont {McGovern}, \citenamefont {Grünzweig},\ and\ \citenamefont {Andersen}}]{McDowall2009}%
  \BibitemOpen
  \bibfield  {author} {\bibinfo {author} {\bibfnamefont {P.}~\bibnamefont {McDowall}}, \bibinfo {author} {\bibfnamefont {A.}~\bibnamefont {Hilliard}}, \bibinfo {author} {\bibfnamefont {M.}~\bibnamefont {McGovern}}, \bibinfo {author} {\bibfnamefont {T.}~\bibnamefont {Grünzweig}},\ and\ \bibinfo {author} {\bibfnamefont {M.~F.}\ \bibnamefont {Andersen}},\ }\bibfield  {title} {\bibinfo {title} {A fidelity treatment of near-resonant states in the atom-optics kicked rotor},\ }\href {https://doi.org/10.1088/1367-2630/11/12/123021} {\bibfield  {journal} {\bibinfo  {journal} {New J. Phys.}\ }\textbf {\bibinfo {volume} {11}},\ \bibinfo {pages} {123021} (\bibinfo {year} {2009})}\BibitemShut {NoStop}%
\bibitem [{\citenamefont {Billam}\ and\ \citenamefont {Gardiner}(2009)}]{Billam2009}%
  \BibitemOpen
  \bibfield  {author} {\bibinfo {author} {\bibfnamefont {T.~P.}\ \bibnamefont {Billam}}\ and\ \bibinfo {author} {\bibfnamefont {S.~A.}\ \bibnamefont {Gardiner}},\ }\bibfield  {title} {\bibinfo {title} {Quantum resonances in an atom-optical $\ensuremath{\delta}$-kicked harmonic oscillator},\ }\href {https://doi.org/10.1103/PhysRevA.80.023414} {\bibfield  {journal} {\bibinfo  {journal} {Phys. Rev. A}\ }\textbf {\bibinfo {volume} {80}},\ \bibinfo {pages} {023414} (\bibinfo {year} {2009})}\BibitemShut {NoStop}%
\bibitem [{\citenamefont {Talukdar}\ \emph {et~al.}(2010)\citenamefont {Talukdar}, \citenamefont {Shrestha},\ and\ \citenamefont {Summy}}]{Talukdar2010}%
  \BibitemOpen
  \bibfield  {author} {\bibinfo {author} {\bibfnamefont {I.}~\bibnamefont {Talukdar}}, \bibinfo {author} {\bibfnamefont {R.}~\bibnamefont {Shrestha}},\ and\ \bibinfo {author} {\bibfnamefont {G.~S.}\ \bibnamefont {Summy}},\ }\bibfield  {title} {\bibinfo {title} {Sub-fourier characteristics of a $\ensuremath{\delta}$-kicked-rotor resonance},\ }\href {https://doi.org/10.1103/PhysRevLett.105.054103} {\bibfield  {journal} {\bibinfo  {journal} {Phys. Rev. Lett.}\ }\textbf {\bibinfo {volume} {105}},\ \bibinfo {pages} {054103} (\bibinfo {year} {2010})}\BibitemShut {NoStop}%
\bibitem [{\citenamefont {Tian}\ and\ \citenamefont {Altland}(2010)}]{Tian2010}%
  \BibitemOpen
  \bibfield  {author} {\bibinfo {author} {\bibfnamefont {C.}~\bibnamefont {Tian}}\ and\ \bibinfo {author} {\bibfnamefont {A.}~\bibnamefont {Altland}},\ }\bibfield  {title} {\bibinfo {title} {Theory of localization and resonance phenomena in the quantum kicked rotor},\ }\href {https://doi.org/10.1088/1367-2630/12/4/043043} {\bibfield  {journal} {\bibinfo  {journal} {New J. Phys.}\ }\textbf {\bibinfo {volume} {12}},\ \bibinfo {pages} {043043} (\bibinfo {year} {2010})}\BibitemShut {NoStop}%
\bibitem [{\citenamefont {Sadgrove}\ \emph {et~al.}(2011)\citenamefont {Sadgrove}, \citenamefont {Wimberger}, \citenamefont {Arimondo}, \citenamefont {Berman},\ and\ \citenamefont {Lin}}]{Sadgrove2011}%
  \BibitemOpen
  \bibfield  {author} {\bibinfo {author} {\bibfnamefont {M.}~\bibnamefont {Sadgrove}}, \bibinfo {author} {\bibfnamefont {S.}~\bibnamefont {Wimberger}}, \bibinfo {author} {\bibfnamefont {E.}~\bibnamefont {Arimondo}}, \bibinfo {author} {\bibfnamefont {P.~R.}\ \bibnamefont {Berman}},\ and\ \bibinfo {author} {\bibfnamefont {C.~C.}\ \bibnamefont {Lin}},\ }\bibfield  {title} {\bibinfo {title} {Chapter 7 - a pseudoclassical method for the atom-optics kicked rotor: from theory to experiment and back},\ }in\ \href {https://doi.org/10.1016/B978-0-12-385508-4.00007-3} {\emph {\bibinfo {booktitle} {Advances In Atomic, Molecular, and Optical Physics}}},\ Vol.~\bibinfo {volume} {60}\ (\bibinfo  {publisher} {Academic Press},\ \bibinfo {year} {2011})\ pp.\ \bibinfo {pages} {315--369}\BibitemShut {NoStop}%
\bibitem [{\citenamefont {Dubertrand}\ \emph {et~al.}(2012)\citenamefont {Dubertrand}, \citenamefont {Guarneri},\ and\ \citenamefont {Wimberger}}]{Dubertrand2012}%
  \BibitemOpen
  \bibfield  {author} {\bibinfo {author} {\bibfnamefont {R.}~\bibnamefont {Dubertrand}}, \bibinfo {author} {\bibfnamefont {I.}~\bibnamefont {Guarneri}},\ and\ \bibinfo {author} {\bibfnamefont {S.}~\bibnamefont {Wimberger}},\ }\bibfield  {title} {\bibinfo {title} {Fidelity for kicked atoms with gravity near a quantum resonance},\ }\href {https://doi.org/10.1103/PhysRevE.85.036205} {\bibfield  {journal} {\bibinfo  {journal} {Phys. Rev. E}\ }\textbf {\bibinfo {volume} {85}},\ \bibinfo {pages} {036205} (\bibinfo {year} {2012})}\BibitemShut {NoStop}%
\bibitem [{\citenamefont {Ullah}\ \emph {et~al.}(2012)\citenamefont {Ullah}, \citenamefont {Ruddell}, \citenamefont {Currivan},\ and\ \citenamefont {Hoogerland}}]{Ullah2012}%
  \BibitemOpen
  \bibfield  {author} {\bibinfo {author} {\bibfnamefont {A.}~\bibnamefont {Ullah}}, \bibinfo {author} {\bibfnamefont {S.~K.}\ \bibnamefont {Ruddell}}, \bibinfo {author} {\bibfnamefont {J.~A.}\ \bibnamefont {Currivan}},\ and\ \bibinfo {author} {\bibfnamefont {M.~D.}\ \bibnamefont {Hoogerland}},\ }\bibfield  {title} {\bibinfo {title} {Quantum resonant effects in the delta-kicked rotor revisited},\ }\href {https://doi.org/10.1140/epjd/e2012-30171-8} {\bibfield  {journal} {\bibinfo  {journal} {Eur. Phys. J. D}\ }\textbf {\bibinfo {volume} {66}},\ \bibinfo {pages} {315} (\bibinfo {year} {2012})}\BibitemShut {NoStop}%
\bibitem [{\citenamefont {Anand}\ \emph {et~al.}(2024)\citenamefont {Anand}, \citenamefont {Davis},\ and\ \citenamefont {Ghose}}]{Anand2024}%
  \BibitemOpen
  \bibfield  {author} {\bibinfo {author} {\bibfnamefont {A.}~\bibnamefont {Anand}}, \bibinfo {author} {\bibfnamefont {J.}~\bibnamefont {Davis}},\ and\ \bibinfo {author} {\bibfnamefont {S.}~\bibnamefont {Ghose}},\ }\bibfield  {title} {\bibinfo {title} {Quantum recurrences in the kicked top},\ }\href {https://doi.org/10.1103/PhysRevResearch.6.023120} {\bibfield  {journal} {\bibinfo  {journal} {Phys. Rev. Res.}\ }\textbf {\bibinfo {volume} {6}},\ \bibinfo {pages} {023120} (\bibinfo {year} {2024})}\BibitemShut {NoStop}%
\bibitem [{\citenamefont {Agarwal}(1981)}]{Agarwal1981}%
  \BibitemOpen
  \bibfield  {author} {\bibinfo {author} {\bibfnamefont {G.~S.}\ \bibnamefont {Agarwal}},\ }\bibfield  {title} {\bibinfo {title} {Relation between atomic coherent-state representation, state multipoles, and generalized phase-space distributions},\ }\href {https://doi.org/10.1103/PhysRevA.24.2889} {\bibfield  {journal} {\bibinfo  {journal} {Phys. Rev. A}\ }\textbf {\bibinfo {volume} {24}},\ \bibinfo {pages} {2889} (\bibinfo {year} {1981})}\BibitemShut {NoStop}%
\bibitem [{sup()}]{sup}%
  \BibitemOpen
  \href@noop {} {}\bibinfo {note} {See Supplemental Material.}\BibitemShut {Stop}%
\bibitem [{\citenamefont {Bonifacio}\ \emph {et~al.}(1971)\citenamefont {Bonifacio}, \citenamefont {Schwendimann},\ and\ \citenamefont {Haake}}]{Bonifacio1971}%
  \BibitemOpen
  \bibfield  {author} {\bibinfo {author} {\bibfnamefont {R.}~\bibnamefont {Bonifacio}}, \bibinfo {author} {\bibfnamefont {P.}~\bibnamefont {Schwendimann}},\ and\ \bibinfo {author} {\bibfnamefont {F.}~\bibnamefont {Haake}},\ }\bibfield  {title} {\bibinfo {title} {Quantum statistical theory of superradiance. i},\ }\href {https://doi.org/10.1103/PhysRevA.4.302} {\bibfield  {journal} {\bibinfo  {journal} {Phys. Rev. A}\ }\textbf {\bibinfo {volume} {4}},\ \bibinfo {pages} {302} (\bibinfo {year} {1971})}\BibitemShut {NoStop}%
\bibitem [{\citenamefont {Gross}\ and\ \citenamefont {Haroche}(1982)}]{Gross1982}%
  \BibitemOpen
  \bibfield  {author} {\bibinfo {author} {\bibfnamefont {M.}~\bibnamefont {Gross}}\ and\ \bibinfo {author} {\bibfnamefont {S.}~\bibnamefont {Haroche}},\ }\bibfield  {title} {\bibinfo {title} {Superradiance: An essay on the theory of collective spontaneous emission},\ }\href {https://doi.org/10.1016/0370-1573(82)90102-8} {\bibfield  {journal} {\bibinfo  {journal} {Phys. Rep.}\ }\textbf {\bibinfo {volume} {93}},\ \bibinfo {pages} {301} (\bibinfo {year} {1982})}\BibitemShut {NoStop}%
\bibitem [{\citenamefont {Haake}\ \emph {et~al.}(2018)\citenamefont {Haake}, \citenamefont {Gnutzmann},\ and\ \citenamefont {Ku{\'{s}}}}]{Haake2018}%
  \BibitemOpen
  \bibfield  {author} {\bibinfo {author} {\bibfnamefont {F.}~\bibnamefont {Haake}}, \bibinfo {author} {\bibfnamefont {S.}~\bibnamefont {Gnutzmann}},\ and\ \bibinfo {author} {\bibfnamefont {M.}~\bibnamefont {Ku{\'{s}}}},\ }\bibinfo {title} {Dissipative systems},\ in\ \href {https://doi.org/10.1007/978-3-319-97580-1_12} {\emph {\bibinfo {booktitle} {Quantum Signatures of Chaos}}}\ (\bibinfo  {publisher} {Springer International Publishing},\ \bibinfo {address} {Cham},\ \bibinfo {year} {2018})\ pp.\ \bibinfo {pages} {591--653}\BibitemShut {NoStop}%
\bibitem [{\citenamefont {Allred}\ \emph {et~al.}(2002)\citenamefont {Allred}, \citenamefont {Lyman}, \citenamefont {Kornack},\ and\ \citenamefont {Romalis}}]{Allred2002}%
  \BibitemOpen
  \bibfield  {author} {\bibinfo {author} {\bibfnamefont {J.~C.}\ \bibnamefont {Allred}}, \bibinfo {author} {\bibfnamefont {R.~N.}\ \bibnamefont {Lyman}}, \bibinfo {author} {\bibfnamefont {T.~W.}\ \bibnamefont {Kornack}},\ and\ \bibinfo {author} {\bibfnamefont {M.~V.}\ \bibnamefont {Romalis}},\ }\bibfield  {title} {\bibinfo {title} {High-sensitivity atomic magnetometer unaffected by spin-exchange relaxation},\ }\href {https://doi.org/10.1103/PhysRevLett.89.130801} {\bibfield  {journal} {\bibinfo  {journal} {Phys. Rev. Lett.}\ }\textbf {\bibinfo {volume} {89}},\ \bibinfo {pages} {130801} (\bibinfo {year} {2002})}\BibitemShut {NoStop}%
\bibitem [{\citenamefont {Kominis}\ \emph {et~al.}(2003)\citenamefont {Kominis}, \citenamefont {Kornack}, \citenamefont {Allred},\ and\ \citenamefont {Romalis}}]{Kominis2003}%
  \BibitemOpen
  \bibfield  {author} {\bibinfo {author} {\bibfnamefont {I.~K.}\ \bibnamefont {Kominis}}, \bibinfo {author} {\bibfnamefont {T.~W.}\ \bibnamefont {Kornack}}, \bibinfo {author} {\bibfnamefont {J.~C.}\ \bibnamefont {Allred}},\ and\ \bibinfo {author} {\bibfnamefont {M.~V.}\ \bibnamefont {Romalis}},\ }\bibfield  {title} {\bibinfo {title} {A subfemtotesla multichannel atomic magnetometer},\ }\href {https://doi.org/10.1038/nature01484} {\bibfield  {journal} {\bibinfo  {journal} {Nature}\ }\textbf {\bibinfo {volume} {422}},\ \bibinfo {pages} {596} (\bibinfo {year} {2003})}\BibitemShut {NoStop}%
\bibitem [{\citenamefont {Savukov}\ and\ \citenamefont {Romalis}(2005)}]{Savukov2005}%
  \BibitemOpen
  \bibfield  {author} {\bibinfo {author} {\bibfnamefont {I.~M.}\ \bibnamefont {Savukov}}\ and\ \bibinfo {author} {\bibfnamefont {M.~V.}\ \bibnamefont {Romalis}},\ }\bibfield  {title} {\bibinfo {title} {Effects of spin-exchange collisions in a high-density alkali-metal vapor in low magnetic fields},\ }\href {https://doi.org/10.1103/PhysRevA.71.023405} {\bibfield  {journal} {\bibinfo  {journal} {Phys. Rev. A}\ }\textbf {\bibinfo {volume} {71}},\ \bibinfo {pages} {023405} (\bibinfo {year} {2005})}\BibitemShut {NoStop}%
\bibitem [{\citenamefont {Budker}\ and\ \citenamefont {Romalis}(2007)}]{Budker2007}%
  \BibitemOpen
  \bibfield  {author} {\bibinfo {author} {\bibfnamefont {D.}~\bibnamefont {Budker}}\ and\ \bibinfo {author} {\bibfnamefont {M.}~\bibnamefont {Romalis}},\ }\bibfield  {title} {\bibinfo {title} {Optical magnetometry},\ }\href {https://doi.org/10.1038/nphys566} {\bibfield  {journal} {\bibinfo  {journal} {Nat. Phys.}\ }\textbf {\bibinfo {volume} {3}},\ \bibinfo {pages} {227} (\bibinfo {year} {2007})}\BibitemShut {NoStop}%
\bibitem [{\citenamefont {Sheng}\ \emph {et~al.}(2013)\citenamefont {Sheng}, \citenamefont {Li}, \citenamefont {Dural},\ and\ \citenamefont {Romalis}}]{Sheng2013}%
  \BibitemOpen
  \bibfield  {author} {\bibinfo {author} {\bibfnamefont {D.}~\bibnamefont {Sheng}}, \bibinfo {author} {\bibfnamefont {S.}~\bibnamefont {Li}}, \bibinfo {author} {\bibfnamefont {N.}~\bibnamefont {Dural}},\ and\ \bibinfo {author} {\bibfnamefont {M.~V.}\ \bibnamefont {Romalis}},\ }\bibfield  {title} {\bibinfo {title} {Subfemtotesla scalar atomic magnetometry using multipass cells},\ }\href {https://doi.org/10.1103/PhysRevLett.110.160802} {\bibfield  {journal} {\bibinfo  {journal} {Phys. Rev. Lett.}\ }\textbf {\bibinfo {volume} {110}},\ \bibinfo {pages} {160802} (\bibinfo {year} {2013})}\BibitemShut {NoStop}%
\end{thebibliography}%

\ifarXiv
    

\section*{Supplementary material}

# Supplementary Material for "Enhancing Quantum Metrology by Quantum Resonance Dynamics"

## I. NUMERICAL APPROACHES FOR DYNAMICS OF QUANTUM FISHER INFORMATION

In our simulations, we calculate the Quantum Fisher Information (QFI) using its relation to the Loschimt echo (fidelity) as follows:

$$I(t) = \lim_{\epsilon \to 0} 4 \frac{1 - F_\epsilon(t)}{\epsilon^2}. \tag{S1}$$

with the Loschimt echo defined as

$$F_\epsilon(t) = |\langle \psi_\alpha(0) | U_\alpha(t) U_{\alpha+\epsilon}(-t) | \psi_\alpha(0) \rangle|^2 \tag{S2}$$

for pure states $|\psi_\alpha\rangle$ we use in the simulations without dissipation. The unitary evolution operator is given by:

$$U_\alpha(t) = \exp(-i\alpha J_z) \exp\left(-i \frac{\beta J_y^2}{2j}\right). \tag{S3}$$

To compute $F_\epsilon(t)$, we perform exact multiplications of the unitary operator iteratively over multiple time steps. To ensure the proper limit in Eq. (S1), we have verified that varying $\epsilon$ within the range $[10^{-9}, 10^{-7}]$ does not affect the results, confirming the robustness of our calculations.

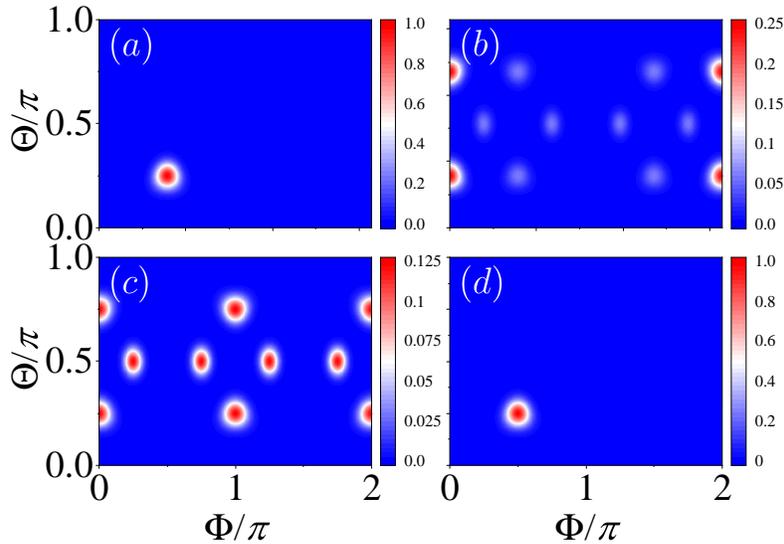

FIG. S1. The Husimi function representation $P(\Theta, \Phi)$ of the wavepacket dynamics of the QKT introduced in the main text, with $\alpha = \frac{\pi}{2}$, $\beta = \pi j/2$, $T = 1$, and $j = 56$ at different evolution times in one period: (a) $t = 0T$, (b) $t = 15T$, (c) $t = 35T$, and (d) $t = 48T$. The quantum recurrence features here enable a highly efficient metrology protocol: with the initial state (a) and the final state (d) staying as a coherent state, the intermediate quantum states (b) and (c) are GHZ-like, highly entangled states potentially yielding a rapid growth in the QFI.

For the dissipative case, where the Loschimit echo should be redefined as

$$F_\epsilon(t) = \left(\text{Tr}\sqrt{[\rho_\alpha^{1/2} \rho_{\alpha+\epsilon}^{1/2}][\rho_\alpha^{1/2} \rho_{\alpha+\epsilon}^{1/2}]^\dagger}\right)^2, \tag{S4}$$

we evolve the system dynamics using the following equation:

$$\rho(t+T) = U_\alpha \left[\exp(\Gamma T)\rho(t)\right] U_\alpha^\dagger, \tag{S5}$$

where the evolution is decomposed into two major steps: 1. The standard unitary evolution by applying $U_\alpha$; 2. The dissipative evolution governed by the operator $\exp(\Gamma T)$ during the interval between two adjacent kicks, $nT \to (n+1)T$. In the dissipative step, the process $\exp(\Gamma T)\rho(t)$ is treated as a dynamical system with $N \times N$ ordinary differential equations. To accurately integrate the dissipative dynamics, we employ a fourth-order Runge-Kutta method. The period of the kicks is fixed at $T = 1$, and the time step for numerical integration is set to $\Delta t = 0.001$ to ensure both accuracy and the stability of our simulation.

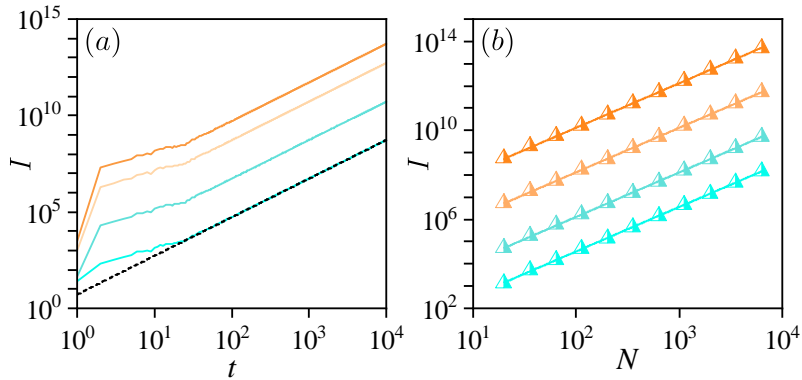

FIG. S2. Dynamics of the QFI associated with the quantum resonances of the QKT. (a) The time evolution of the QFI. Cyan to orange curves represent different system sizes ranging from $N = 20$ to $N = 6324$, with $\beta = \pi j/2$. The black dashed line represents a power-law fit, showing $I \sim t^a$ with $a \approx 2$. (b) The dependence of the QFI on the number of spins for $\beta = \pi j$ at different times $t/T = 10, 10^2, 10^3, 10^4$ (from cyan to orange). The solid lines represent a fitting with $I \sim N^a$, yielding $a \approx 2$, demonstrating the Heisenberg limit scaling.

## II. QUANTUM RECURRENCE IN THE QUANTUM KICKED TOP UNDER PARAMETER $\beta = j\pi/2$

Here, we present the quantum dynamics of the quantum kicked top for the case $\beta = j\pi/2$, which exhibits complete quantum recurrence with a period of $48T$ (case (iii) discussed in the main text). This parameter choice leads to the generation of multiple GHZ-like states during the evolution while preserving the recurrence of the coherent state, as shown in Fig. S1. These unique dynamical features make this parameter setting a particularly promising candidate for quantum metrology.

Similar to case (ii), the enhanced scaling of the QFI is also observed in this case, as shown in Fig. S2. This serves as a supplementary analysis to our discussion of case (ii) in the main text.

## III. STABILITY OF THE QUANTUM RECURRENCE DYNAMICS UNDER PERTURBATIONS

In the main text, we demonstrate that under the near-resonance condition $\beta = j\pi + \delta$, the Heisenberg limit scaling $I(t) \propto N^2 t^2$ is preserved. Here, we show that while quantum recurrence is no longer perfect under such perturbations, the essential physical mechanism enabling this advanced scaling remains intact. Specifically, the intermediate states continue to exhibit GHZ-like characteristics, albeit with some distortion, as presentd in Fig. S3 for $\delta = 1$.

## IV. ANALYTICAL RESULTS OF FISHER INFORMATION FOR $\beta = \pi j$

In this section, we derive the exact quantum recurrence dynamics under the resonance condition $\beta = \pi j$. Consider an initial coherent state $|\psi(0)\rangle = |\Theta, \Phi\rangle$, which can be expressed it in the eigenbasis $|m\rangle$ of the operator $J_y$ as

$$|\Theta, \Phi\rangle = \sum_m c_m |m\rangle. \tag{S6}$$

Note that coherent state $|\pi - \Theta, \pi - \Phi\rangle$ can be related to $|\Theta, \Phi\rangle$ through an unitary operation $\exp(-i\pi J_y)$

$$|\pi - \Theta, \pi - \Phi\rangle = \exp(-i\pi J_y)|\Theta, \Phi\rangle = \sum_{\text{mod }(m,2)=0} c_m |m\rangle - \sum_{\text{mod }(m,2)=1} c_m |m\rangle. \tag{S7}$$

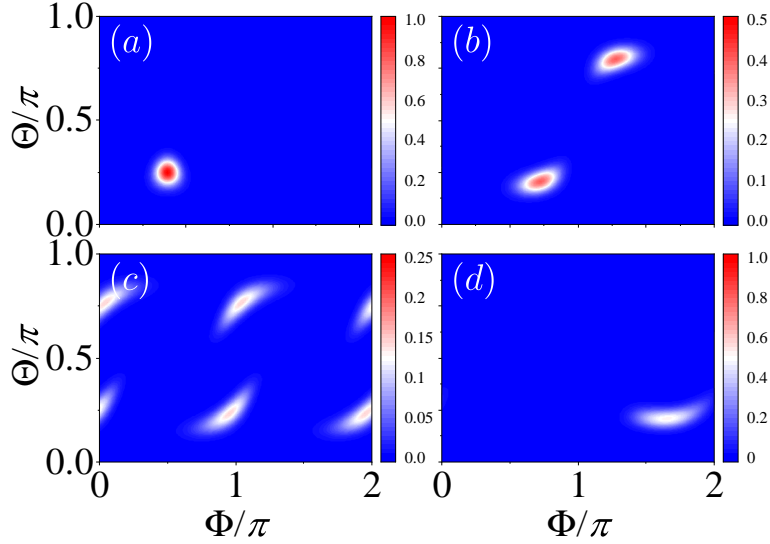

FIG. S3. The Husimi function representation $P(\Theta, \Phi)$ of the wavepacket dynamics of the QKT introduced in the main text, with $\alpha = \frac{\pi}{2}$, $\beta = \pi j + 1$, $T = 1$, and $j = 56$ at different evolution times in one period: (a) $t = 0T$, (b) $t = 3T$, (c) $t = 6T$, and (d) $t = 8T$. The quantum recurrence features here enable a highly efficient metrology protocol: with the initial state (a) and the final state (d) staying as a coherent state, the intermediate quantum states (b) and (c) are GHZ-like, highly entangled states potentially yielding a rapid growth in the QFI.

Based on Eqs. (S6) and (S7), the result after applying evolution operator $U$ to $|\Theta, \Phi\rangle$ can be expressed as

$$\exp\left(-i\frac{\pi}{2}J_z\right)\exp\left(-i\frac{\pi}{2}J_y^2\right)|\Theta, \Phi\rangle = \exp(-i\frac{\pi}{2}J_z)\left(\sum_{\text{mod }(m,2)=0} c_m|m\rangle - i\sum_{\text{mod }(m,2)=1} c_m|m\rangle\right)$$
$$= \exp(-i\frac{\pi}{2}J_z)\left(\frac{1-i}{2}|\Theta, \Phi\rangle + \frac{1+i}{2}|\pi-\Theta, \pi-\Phi\rangle\right) \qquad (S8)$$
$$= \frac{1-i}{2}\left|\Theta, \Phi + \frac{\pi}{2}\right\rangle + \frac{1+i}{2}\left|\pi-\Theta, \frac{3\pi}{2}-\Phi\right\rangle.$$

After some straightforward calculations, the wavefunction at time $t$ can be expressed as

$$|\psi(t)\rangle = \begin{cases} |\Theta, \Phi\rangle,, & \text{if } \mod(t,8) = 0, \\ \frac{1-i}{2}\left|\Theta, \Phi+\frac{\pi}{2}\right\rangle + \frac{1+i}{2}\left|\pi-\Theta, \frac{3\pi}{2}-\Phi\right\rangle, & \text{if } \mod(t,8) = 1, \\ -\frac{i}{2}|\Theta, \Phi+\pi\rangle + \frac{1}{2}|\pi-\Theta, \pi-\Phi\rangle + \frac{1}{2}|\pi-\Theta, 2\pi-\Phi\rangle + \frac{i}{2}|\Theta, \Phi\rangle, & \text{if } \mod(t,8) = 2, \\ \frac{1-i}{2}\left|\pi-\Theta, \frac{\pi}{2}-\Phi\right\rangle + \frac{1+i}{2}\left|\Theta, \Phi+\frac{\pi}{2}\right\rangle, & \text{if } \mod(t,8) = 3, \\ |\Theta, \Phi+\pi\rangle, & \text{if } \mod(t,8) = 4, \\ \frac{1-i}{2}\left|\Theta, \Phi+\frac{3\pi}{2}\right\rangle + \frac{1+i}{2}\left|\pi-\Theta, \frac{\pi}{2}-\Phi\right\rangle, & \text{if } \mod(t,8) = 5, \\ -\frac{i}{2}|\Theta, \Phi\rangle + \frac{1}{2}|\pi-\Theta, 2\pi-\Phi\rangle + \frac{1}{2}|\pi-\Theta, \pi-\Phi\rangle + \frac{i}{2}|\Theta, \Phi+\pi\rangle, & \text{if } \mod(t,8) = 6, \\ \frac{1-i}{2}\left|\pi-\Theta, \frac{3\pi}{2}-\Phi\right\rangle + \frac{1+i}{2}\left|\Theta, \Phi+\frac{3\pi}{2}\right\rangle, & \text{if } \mod(t,8) = 7. \end{cases} \qquad (S9)$$

According to the definition of quantum fisher information, the information about parameter $\alpha$ in $|\psi(t)\rangle$ is

$$I(t) = 4\left[\langle\partial_\alpha\psi(t)|\partial_\alpha\psi(t)\rangle - |\langle\psi(t)|\partial_\alpha\psi(t)\rangle|^2\right], \qquad (S10)$$

where $|\psi(t)\rangle = U^t|\psi(0)\rangle$ is the wavefunction at time $t$. Here the derivative of wavefunction on $\alpha$ at time $t$ can be expressed as

$$|\partial_\alpha\psi(t)\rangle = \left[U^t\left(-iJ_z\right) + U^{t-1}\left(-iJ_z\right)U + \cdots + U\left(-iJ_z\right)U^{t-1}\right]|\psi(0)\rangle. \qquad (S11)$$

Therefore the inner product of $|\psi(t)\rangle$ and $|\partial_\alpha\psi(t)\rangle$ is

$$\langle\psi(t)|\partial_\alpha\psi(t)\rangle = \sum_{i=0}^{t-1}\langle\psi(i)|-iJ_z|\psi(i)\rangle. \qquad (S12)$$

Due to the periodic dynamics, above result can be calculated analytically as

$$\langle\psi(t)|\partial_\alpha\psi(t)\rangle = -\mathrm{i}\left\lceil\frac{t}{4}\right\rceil j\cos\Theta. \tag{S13}$$

For the first term in Eq. (S10), due to the fact that the overlap between two coherent with different center is small, we can approximate it as

$$\begin{aligned}
\langle\partial_\alpha\psi(t)|\partial_\alpha\psi(t)\rangle &= \sum_{i=0}^{t-1}\sum_{j=0}^{t-1}\langle\psi(i)|J_z U^{i-j} J_z|\psi(j)\rangle \\
&\approx \sum_{i=0}^{t-1}\sum_{j=0}^{t-1}\langle\psi(i)|J_z^2|\psi(j)\rangle\delta_{\mathrm{mod}\,(i-j,8),0} \\
&= \left(j^2\cos^2\Theta + \frac{j}{2}\sin^2\Theta\right)\sum_{i=0}^{t-1}\sum_{j=0}^{t-1}\delta_{\mathrm{mod}\,(i-j,8),0} \\
&= \left(j^2\cos^2\Theta + \frac{j}{2}\sin^2\Theta\right)\left((8-\mathrm{mod}\,(t,8))\left\lfloor\frac{t}{8}\right\rfloor^2 + \mathrm{mod}\,(t,8)\left\lceil\frac{t}{8}\right\rceil^2\right).
\end{aligned} \tag{S14}$$

Substitute Eq.(S13) and Eq. (S14) into Eq. (S10), the final result of fisher information is

$$I(t) = 4\left[\left(j^2\cos^2\Theta + \frac{j}{2}\sin^2\Theta\right)\left((8-\mathrm{mod}\,(t,8))\left\lfloor\frac{t}{8}\right\rfloor^2 + \mathrm{mod}\,(t,8)\left\lceil\frac{t}{8}\right\rceil^2\right) - j^2\cos^2\Theta\left\lceil\frac{t}{4}\right\rceil^2\right]. \tag{S15}$$

Consider a special time $t = 8mT$ $(m \in \mathbb{Z})$, above result can be simplified to

$$I(t) = 16\left(j^2\cos^2\Theta + j\sin^2\Theta\right)m^2. \tag{S16}$$


\fi

\end{document}